\documentclass[a4paper,oneside,final,notitlepage,onecolumn]{article}
\usepackage{amsfonts}
\usepackage{epsf}
\usepackage{graphicx}% Include figure files
\usepackage{amssymb,eso-pic}
\usepackage{latexsym}
\usepackage{tabularx}
\usepackage{amsxtra}
\usepackage{hyperref}
\usepackage{t1enc}
\usepackage[normalem]{ulem}
\usepackage{color}
\definecolor{blue}{rgb}{0,0,1}
\definecolor{red}{rgb}{1,0,0}

\DeclareFontFamily{OT1}{rsfs}{} \DeclareFontShape{OT1}{rsfs}{m}{n}{
<-7> rsfs5 <7-10> rsfs7 <10-> rsfs10}{}
\DeclareMathAlphabet{\mycal}{OT1}{rsfs}{m}{n}

\def\sc{{\hskip 3.5pt {{}^{{}^{{}_{{}_{\bowtie}}}}} \kern -8.pt{}}}  
\def\SC{{\hskip 3.5pt {{}^{{}^{{}^{{}_{{}_{\bowtie}}}}}} \kern -10.5pt{}}}

\def\tf{\textit{\texttt{f}}}
\def\tz{\textit{\texttt{z}}}

\begin{document}

%%%%%%%%%%%%%%%%%%%%%%%%%%%%%%%%%%%%%%%%%%%%%%%%%%%%%%%%%%%%%%%%%%%%%%%%%%%%%%
\newtheorem{theorem}{Theorem}[section]
\newtheorem{lemma}{Lemma}[section]
\newtheorem{proposition}{Proposition}[section]
\newtheorem{corollary}{Corollary}[section]
\newtheorem{conjecture}{Conjecture}[section]
\newtheorem{example}{Example}[section]
\newtheorem{definition}{Definition}[section]
\newtheorem{remark}{Remark}[section]
\newtheorem{exercise}{Exercise}[section]
\newtheorem{axiom}{Axiom}[section]
%%%%%%%%%%%%%%%%%%%%%%%%%%%%%%%%%%%%%%%%%%%%%%%%%%%%%%%%%%%%%%%%%%%%%%%%%%%%%%
\renewcommand{\theequation}{\thesection.\arabic{equation}} 
% A fenti parancs atdefinialja az egyenleteket szamozo parancsot
%%%%%%%%%%%%%%%%%%%%%%%%%%%%%%%%%%%%%%%%%%%%%%%%%%%%%%%%%%%%%%%%%%%%%%%%%%%%%%

\author{ Istv\'an R\'{a}cz\thanks{% 
~email: racz.istvan@wigner.mta.hu}  \\ %EndAName 
Wigner RCP, \\ H-1121
Budapest, Konkoly Thege Mikl\'os \'ut 29-33. \\Hungary
}
\title{Stationary Black Holes as Holographs II.}  
\maketitle

\begin{abstract}

{Within the generic null characteristic initial value problem a reduced set of the evolution equations are deduced from the coupled Newman-Penrose and Maxwell equations for} smooth four-dimensional electrovacuum spacetimes {allowing non-zero cosmological constant}. {It is shown that this reduced equations make up a first order symmetric hyperbolic system of evolution equations, and also that the solutions to this reduced system  are also solutions to the full set of the Newman-Penrose and Maxwell equations provided that the {\it inner} equations hold on the initial data surfaces. The derived generic results} are {applied in carrying out the investigation of electrovacuum spacetimes distinguished by the existence of} a pair of null hypersurfaces, $\mathcal{H}_1$ and $\mathcal{H}_2$, generated by expansion and shear free geodesically complete null congruences such that they intersect on a two-dimensional 
spacelike surface, $\mathcal{Z}=\mathcal{H}_1\cap\mathcal{H}_2$. Besides the existence of this pair of null hypersurfaces no assumption concerning the asymptotic structure is made. {I}t is shown that both the spacetime geometry and the electromagnetic field are uniquely determined, in 
the domain of dependence of $\mathcal{H}_1\cup\mathcal{H}_2$ once a complex vector field $\xi^A$ (determining the metric induced on $\mathcal{Z}$), the $\tau$ spin coefficient and the $\phi_1$ electromagnetic potential are specified on $\mathcal{Z}$.
The existence of a Killing vector field---with respect to which the null hypersurfaces $\mathcal{H}_1$ and $\mathcal{H}_2$ comprise a bifurcate type Killing horizon---is also justified in the domain of dependence of $\mathcal{H}_1\cup\mathcal{H}_2$. Since, in general, the freely specifiable data on $\mathcal{Z}$ do not have any sort of symmetry the corresponding spacetimes do not possess any symmetry in addition to the horizon Killing vector field. Thereby, they comprise the class of generic `stationary' distorted electrovacuum black hole spacetimes {which for the case of positive cosmological constant may also (or, in certain cases, only) contain a distorted de Sitter type cosmological horizon to which our results equally apply}. 
It is also shown that there are stationary distorted electrovacuum black hole configurations such that parallelly propagated curvature blow up occurs both to the future and to the past ends of some of the null generators of their bifurcate Killing horizon, and also that this behavior is universal. 
In particular, it is shown that, in the space of vacuum solutions to Einstein's equations, in an arbitrarily small neighborhood of the Schwarzschild solution this type of distorted vacuum black hole configurations always exist. A short discussion on the relation of these results and some of the recent claims on the instability of extremal black holes is also given.  
\end{abstract}

\section{Introduction}
\setcounter{equation}{0}

{Since the characteristic initial value problem is expected to provide  an adequate and powerful setup to study gravitational wave propagation in the full complexity of Einstein's theory form the early 60's (see, e.g.} \cite{bondi,sachs,penrose,MZH,friedrich,friedrich1,rendall,damour,hayward,chr}{) there have been continuous interest and considerable developments in this field}\,\footnote{{For an excellent recent review on the various formulations and many related issues see \cite{chr}.}} {, therefore, most of these studies---with few exceptions} \cite{rendall,damour}{---were restricted to the vacuum problem. In this respect the analytic setup applied in the present paper is simply one of the possible choices. The specific choice we made was motivated by that of Friedrich used in \cite{friedrich} studying the regular and asymptotic characteristic initial value problem in the vacuum case.  Accordingly, one could conclude that the generic part of the argument of the 
present paper is simply a repetition of results derived elsewhere. To provide some evidences in order to see that this is not so recall first that within the characteristic initial value problem there is a huge freedom even in choosing the freely specifiable and constrained part of the initial data (for more details see \cite{chr}). The choice we made concerning this part of the initial data is again an adaptation of the one used by Friedrich in \cite{friedrich}. This, in particular, means that our choice differs from the ones applied in }\cite{rendall,damour}{, in studying the non-vacuum configurations. There is, however, an even more important additional motivation for the inclusion of the generic argument by making use of the Newman-Penrose variables for the Einstein-Maxwell problem in this paper. It is rooted on remarks made by Chandrasekhar when discussing the role of the full set of the Newman-Penrose equations in Sections 7 and 8 of Chapter 1 in his brilliant book \cite{CH} on black hole 
physics. He made there the following comments: ``It is not clear how many of these equations are independent, how they are to be ordered or used and, indeed, what they are for.'' The generic part of the argument is to provide answers to these implicit questions by making use of a suitable adaptation and slight generalization of Friedrich's results \cite{friedrich} to the non-vacuum problem.}
 
\medskip

{According to the above discussion} the scope of this paper is twofold. Besides providing a systematic investigation of the generic characteristic initial value problem within the setup of Newman-Penrose formalism, using these generic results, a characterization of the full set of the stationary distorted electrovacuum black hole configurations is also given.
 
\medskip

{Originally the distorted black hole solutions were considered to be relevant only in context of the deformations of the well-known four-dimensional asymptotically flat, or asymptotically (locally) anti-de-Sitter, black hole configurations} \cite{israel3,szekeres,GH,FK}{. The deformations were assumed to be exerted by certain hypothetical external mass distributions. Recently it was realized that generic static distorted black hole spacetimes do also play important role in context of higher dimensional vacuum black hole configurations whenever some of the spatial dimensions are compactified (see e.g.}\,\cite{ME,HO,FR2}{).}
 
\medskip

In modeling generic stationary distorted black hole spacetimes it is assumed that they possess only a single (global) one-parameter group of isometries associated with a Killing horizon (see, e.g., \cite{holographs1}). In general, no apriory assumption concerning the asymptotic properties is applied. Accordingly, the class of stationary distorted electrovacuum black hole spacetimes is expected to contain, besides the members of the Kerr-Newman family, a lot of nearby configurations. Interestingly only very limited subclasses of these stationary distorted black hole spacetimes have been investigated. Among these one can find the {\it static axially symmetric} distorted {\it vacuum} black hole configurations, which have been discovered and studied for long (see, e.g.\,\cite{israel3,szekeres,PE,GH,CH,ME,FR1,HO,FR2,FK,Tom,FR3}), and a small subset of the {\it static axisymmetric} distorted {\it electrovacuum} black hole configurations \cite{FK,AS}. 
Therefore, it seems to be of obvious interest to provide a clear characterization of the space of stationary distorted electrovacuum black hole solutions. 

\medskip

Motivated by the above outlined necessities, in the present paper four-dimensional electrovacuum spacetimes in Einstein's theory will be considered which admit a pair of null hypersurfaces, $\mathcal{H}_1$ and $\mathcal{H}_2$, generated by expansion and shear free geodesically complete null congruences such that they intersect on a two-dimensional spacelike surface, $\mathcal{Z}=\mathcal{H}_1\cap\mathcal{H}_2$. 
The true physical degrees of freedom of these generic four-dimensional electrovacuum spacetimes are explored by making use of a combination of the Newman-Penrose formalism and that of the null characteristic initial value problem. It is shown that the geometry and the electromagnetic field are uniquely determined, in the domain of dependence of $\mathcal{H}_1\cup\mathcal{H}_2$, once a complex vector field $\xi^A$---which can be associated with the negative definite metric induced on $\mathcal{Z}$ as $g^{AB}=-(\xi^A\overline\xi^B+\overline\xi^A\xi^B)$---the $\tau$ spin coefficient and the middle electromagnetic potential $\phi_1$ are specified on $\mathcal{Z}$.\,\footnote{{In virtue of the title chosen this paper is a continuation of \cite{holographs1}. Our original intention was to write only a short corrigendum as we have found Lemma 5.2 in \cite{holographs1} to be erroneous. More precisely, the very last equality in relation (5.19) of \cite{holographs1} does not hold. Accordingly the $\tau$ spin 
coefficient need not to vanish on the bifurcation surface. This could make the arguments in  \cite{holographs1} to be pointless as Lemma 5.2 plays central role in deriving several results in \cite{holographs1}. Note, however, that the situation is not as severe as it looks. As it is justified by the main results of the present paper, by assuming the vanishing of the spin coefficient $\tau$ on the bifurcation surface in \cite{holographs1} we simply restricted considerations to a subset of generic `stationary' distorted black hole spacetimes. In other words, the results derived in \cite{holographs1} can be seen to hold on a proper subset of stationary distorted black hole 
configurations.} 

{In order to avoid any sort of uncertainties in the present paper a new self-contained argument is provided covering the electrovacuum case, with allowing non-zero cosmological constant. This is done such that the assumptions are reduced to the minimal set which should yield a clear identification of both the selected class of spacetimes and the applied techniques.}}

In addition, based on results covered by \cite{frw,r1,rkill,rkill2}, the existence of a (global) one-parameter group of isometries in the domain of dependence of $\mathcal{H}_1\cup\mathcal{H}_2$ is also proved. The null hypersurfaces $\mathcal{H}_1$ and $\mathcal{H}_2$ are shown to comprise a bifurcate type Killing horizon with respect to the associated Killing vector field. 
Since, in the generic case, the freely specifiable data on $\mathcal{Z}$ do not have any symmetry the selected class of spacetimes do not either possess any symmetry in addition to the horizon Killing vector field. Besides the existence of this pair of null hypersurfaces no assumption concerning the asymptotic structure is made. Thereby, the selected spacetimes do really comprise the class of generic stationary electrovacuum distorted black hole spacetimes {which for the case of positive cosmological constant may also (or, in certain cases, only) contain a distorted de Sitter type cosmological horizon} such that their event {or cosmological} horizons are bifurcate type Killing horizons comprised by the null hypersurfaces $\mathcal{H}_1$ and $\mathcal{H}_2$.\,\footnote{{Thanks are due to Akihiro Ishibashi for raising questions in context of cosmological horizons to which case our results equally apply as in their derivation, besides the existence of this pair of intersecting null 
hypersurfaces, no assumption concerning the asymptotic structure is made.}} 

\medskip

We have kept referring to the hypersurfaces $\mathcal{H}_1$ and $\mathcal{H}_2$ as if they were null even in context of the characteristic initial value problem. This was done despite that it is not obvious at all in advance of obtaining a solution whether the initial data surfaces can be considered as characteristic surfaces or not. {In this respect it is important to recall that by using the freely specifiable part of the initial data, and solving the constraint equations---these are usually referred as the ``inner equations'' and they are ordinary differential equations along the generators of the transversely intersecting hypersurfaces $\mathcal{H}_1$ and $\mathcal{H}_2$---the metric can always be recovered on $\mathcal{H}_1\cup\mathcal{H}_2$. For a manifestation of the corresponding process see the discussion above Lemma\,\ref{lHF2} and also Section\,\ref{app} of the present paper.} 

{It is also important to know what is the extent of the domain of dependence}\,\footnote{{In a time oriented spacetime consider a pair of transversely intersecting characteristic hypersurfaces $\mathcal{H}_1$ and $\mathcal{H}_2$. Denote by $\mathcal{H}_1^+$, $\mathcal{H}_2^+$ and $\mathcal{H}_1^-$, $\mathcal{H}_2^-$ the parts of $\mathcal{H}_1$ and $\mathcal{H}_2$ which belong to the causal future and past of $\mathcal{Z}$, respectively. Then, the union of the domain of dependencies $D[\mathcal{H}_1^+\cup\mathcal{H}_2^+]$ and $D[\mathcal{H}_1^-\cup\mathcal{H}_2^-]$ of the pairwise achronal hypersurfaces $\mathcal{H}_1^+\cup\mathcal{H}_2^+$ and $\mathcal{H}_1^-\cup\mathcal{H}_2^-$ will be referred to as the domain of dependence  of $\mathcal{H}_1$ and $\mathcal{H}_2$, and it will be denoted by $\mathcal{D}[\mathcal{H}_1\cup\mathcal{H}_2]$. }}, {$\mathcal{D}[\mathcal{H}_1\cup\mathcal{H}_2]$, of $\mathcal{H}_1\cup\mathcal{H}_2$. In this respect it should be emphasized that we shall 
make use of K\'ann\'ar's argument (see Section 3 in \cite{kannar}), which is based on the remarkable 
argument given by Rendall in \cite{rendall}. As both of these arguments guarantee the existence of a unique solution to the characteristic initial value problem only in a sufficiently small neighborhood of the intersection $\mathcal{Z}=\mathcal{H}_1\cap\mathcal{H}_2$ our results (see Theorem\,\ref{unique2}, in particular the last for paragraphs of its proof) also have this limitation, i.e.\,the domain of dependence is restricted to such a neighborhood.}

Note that if considerations were restricted to the vacuum case we could refer to a stronger result. Namely, it was shown recently by Luk \cite{luk} that the solution is guaranteed then to exist in a neighborhood of the hypersurfaces $\mathcal{H}_1$ and $\mathcal{H}_2$, as long as the inner equations are initially satisfied on them. It seems to be plausible that an analogous result applies to the case of electrovacuum spacetimes with the inclusion of non-zero cosmological constant, nevertheless, the investigation of this issue is out of the scope of the present paper.

\medskip

Although the unique determinacy of the four-dimensional electrovacuum configurations by the initial data specified on their bifurcation surface is proved we have to admit that in this paper, besides the well-known Schwarzschild solution,\,\footnote{{As the global existence problem is left out from the discussions of the present paper the domain of dependence pertinent for the Schwarzschild or Kerr-Newman solutions is restricted to a neighborhood of the bifurcation surface just like in the generic case. Note, however, that if the global existence of a three parameter shear free congruence of null geodesics was guaranteed the corresponding particular solutions could also be recovered in a global manner by a suitable adaptation of the local argument applied in the proof of Lemma\,\ref{schwl} below. Nevertheless, as in the present context the existence of such a congruence is verified by Lemma\,\ref{schwl} only locally the domain of dependencies have to be local even for the Schwarzschild or Kerr-Newman 
solutions.}} no fully explicit four-dimensional solutions to the Einstein-Maxwell equations will be given. Nevertheless, all the solutions belonging to the selected class of spacetimes will become explicitly determined on the pair of null hypersurfaces, $\mathcal{H}_1$ and $\mathcal{H}_2$. This way an interesting and unexpected property of these  four-dimensional stationary distorted electrovacuum black hole configurations could be discovered. The occurrence of parallelly propagated curvature blow up, both to the future and to the past ends of some of the null generators of the bifurcate Killing horizon, was found to be universal. 

Based on this universality the following remarkable result could also be justified. Consider the space of four-dimensional vacuum solutions to Einstein's equations. Then, in an arbitrarily small neighborhood of the Schwarzschild solution there always exist distorted vacuum black hole configurations such that parallelly propagated curvature blow up occurs both to the future and to the past ends of some of the null generators of their bifurcate Killing horizon. 

\medskip

This paper is organized as follows: 
Section\,\ref{stac} is to select the class of spacetimes to which our main results apply and to recall some of the basic notions and variables we shall use. Section\,\ref{nchfr} is to set-up the characteristic initial value problem in the generic context by making use of the Newman-Penrose formalism. The proof of the key results is presented separately in Appendix \ref{Appendix A}. 
In Section\,\ref{app} these results are applied to characterize generic four-dimensional stationary distorted electrovacuum black hole spacetimes. 
Section\,\ref{sing} is to show that a parallelly propagated curvature blow up occurs both to the future and past ends of some of the null generators of the event horizons of generic distorted black hole spacetimes. 
Section\,\ref{sym} is to justify the existence of the horizon compatible spacetime symmetry. 
Section\,\ref{sing2} is to demonstrate that in arbitrarily small neighborhood of the Schwarzschild solution there always exist vacuum configurations such that a parallelly propagated curvature blow up occurs at the future and past ends of some of the null generators of the event horizons.
Section\,\ref{con} contains our final remarks, along with a short discussion on the relation of the presented results and some of the recent arguments on the instability of extremal black holes.

\section{Preliminaries}\label{stac}
\setcounter{equation}{0}

Throughout this paper a spacetime $(M,g_{ab})$ is taken to be a four-dimensional smooth, paracompact, connected, orientable manifold $M$ endowed with a smooth Lorentzian metric $g_{ab}$ of signature $(+,-,\dots,-)$. It is
assumed that $(M,g_{ab})$ is  time orientable and that a time
orientation has been chosen.

The electromagnetic field is represented by a $2$-form field $F_{\,ab}$ satisfying the source free Maxwell equations\,\footnote{Here $\nabla_a$ denote the metric compatible Levi-Civita connection with curvature tensor $(\nabla_a \nabla_b - \nabla_b \nabla_a) X_c= X_d {R^d}_{cab}$ as defined in \cite{newman:penrose,friedrich}. Notice that this is opposite to the curvature convention applied in \cite{HE,wald}.}
\begin{equation}\label{m1}
\nabla^aF_{\,ab}=0\ \ \ {\rm and}\ \ \ \nabla_{[a}{F}_{\,b c]}=0\,.
\end{equation}

The metric $g_{ab}$ will be assumed to satisfy the Einstein equations\,\footnote{Note that because of the curvature convention applied here the Ricci and Einstein tensors are minus one times that of the corresponding objects in \cite{HE,wald} which explains the use of the sign factor on the right hand side of (\ref{ein}) and in the definition of the energy momentum tensor.}
\begin{equation}\label{ein}
R_{ab}-\frac{1}{2} g_{ab} R+\widetilde\Lambda g_{ab}=-8\pi T_{ab},
\end{equation}
with cosmological constant $\widetilde\Lambda$, and with the energy momentum tensor\,\footnote{Note that our curvature convention and choice of signature are combined such that the asymptotically anti-de Sitter type configurations do correspond to negative cosmological constant as in \cite{HE,wald}.}  
\begin{equation}\label{max}
T_{ab}=-\frac{1}{4\pi}\left[ F_{\,ae}{F_{\,b}}^e -
\frac{1}{4} g_{ab} \left(F_{\,ef}F^{\,ef}\right) \right]\,.
\end{equation} %There has been here a minus sign as the signature of the metric is opposite to the standard one. however this change is compensated by a factor of -1 coming from the fact that the convention applied in defining the curvature tensor in NP differs from the convention applied in Wald's book. For the very same reason the sign of the scalar curvature in the present case is the same as it is in Wald's book. {R_{abc}}^d=-{R^{NP}_{abc}}^d ===> R_{ab}=-R^{NP}_{ab} & g^{ab}=-g_{NP}^{ab} ===> R=R_{ab}g^{ab}=(-R^{NP}_{ab})(-g_{NP}^{ab})=R^{NP}.....

\medskip

{In deriving our results in context of the generic characteristic initial value problem only the above requirements will be applied. In setting up Gaussian null coordinate systems and the Newman-Penrose formalism, as it is described in the succeeding subsections, a pair of null hypersurfaces $\mathcal{H}_1$ and $\mathcal{H}_2$, which individually possesses the topology $\mathbb{R}\times \mathcal{Z}$ and which intersect on a two-dimensional spacelike surface $\mathcal{Z}=\mathcal{H}_1\cap\mathcal{H}_2$, will also be applied. In characterizing the distorted black hole} spacetime{s, in addition, it will also be assumed that the null geodesic congruences generating these}  null hypersurfaces {are} expansion and shear free {and also that they are} geodesically complete. 

\subsection{Gaussian null coordinates}\label{pre}

Let $k^a$ be a smooth {\it future directed} null vector field on $\mathcal{Z}$  tangent to $\mathcal{H}_2$ and extend it to  $\mathcal{H}_2$ by requiring the relation  $k^e\nabla_e k^a=0$ to hold on ${\mathcal{H}_2}$. Choose $u$ to be affine parameter along the null generators of $\mathcal{H}_2$ synchronized such that $u=0$ on $\mathcal{Z}$. Denote by $\chi_u$ the $1$-parameter family of diffeomorphisms associated with $k^a$ and by $\mathcal{Z}_u=\chi_u[\mathcal{Z}]$ the associated $1$-parameter family of smooth cross-sections of $\mathcal{H}_2$. Choose $\ell^a$ to be the unique {\it future directed} null vector field on $\mathcal{H}_2$ which is orthogonal to the two-dimensional cross-sections $\mathcal{Z}_u$ and satisfies the normalizing condition $\ell^a k_a=1$ on $\mathcal{H}_2$. Consider now the null geodesics starting at the points of $\mathcal{H}_2$ with tangent $\ell^a$. Since $\mathcal{H}_2$ was assumed to be smooth and the vector fields $k^a$ and $\ell^a$ are by construction 
smooth on $\mathcal{H}_2$, these geodesics do not intersect in a sufficiently small open neighborhood 
$\mathcal{O} \subset M$ of $\mathcal{H}_2$. By choosing $r$ to be the affine parameter along the null geodesics starting at the points of $\mathcal{H}_2$ with tangent $\ell^a$ and synchronised so that $r=0$ on $\mathcal{H}_2$ we get a smooth real function $r:\mathcal{O}\rightarrow \mathbb{R}$.  Extend then, the function $u:\mathcal{H}_2\rightarrow \mathbb{R}$ onto $\mathcal{O}$ by requiring its value to be constant along the null geodesics with tangent $\ell^a=\left(\partial/\partial r\right)^a$. The yielded function $u:\mathcal{O}\rightarrow \mathbb{R}$ is by construction smooth. 

\medskip

By choosing suitable local coordinates on patches of $\mathcal{Z}$  Gaussian null coordinates $(u,r,x^3,x^4)$ can be defined everywhere on the corresponding sub-domains of $\mathcal{O}$. In either of these coordinate domains the most general form of the metric can be given as \cite{rw1,rw2,ext2} 
\begin{equation}
\mathrm{d}s^2=g_{uu}\mathrm{d}u^2+2\mathrm{d}r\mathrm{d}u+2 g_{uA}\mathrm{d}u \mathrm{d}x^A+g_{AB}\mathrm{d}x^A\mathrm{d}x^B,
\label{le1}
\end{equation}
where $g_{uu}$, $g_{uA}$ and $g_{AB}$ are smooth functions of the coordinates $u,r,x^3,x^4$ such that $g_{uu}$ and $g_{uA}$ vanish on $\mathcal{H}_2$, and $g_{AB}$ is a negative definite $2\times 2$ matrix. The uppercase Latin indices take the values $3,4$. 

Note that the null hypersurface $\mathcal{H}_1$ can be given as $u=0$, while $\mathcal{H}_2$ does correspond to the $r=0$ null hypersurface. 
It is also worth keeping in mind that while the functions $u$ and $r$ are defined throughout $\mathcal{O}$ the spatial coordinates $x^3,x^4$ may, in general, only be defined in sub-domains of $\mathcal{O}$. By patching these sub-domains---based on the paracompactness of $M$, and thereby that on that of $\mathcal{Z}$ and using an argument analogous to the one covered by the 3-6 paragraphs on page 961 of \cite{kannar}---results that can be derived, in the succeeding sections with the use of the characteristic initial value problem, on either of these sub-domains can always be seen to extend onto the entire of $\mathcal{O}$. Therefore, hereafter, for the sake of simplicity, we will keep referring to $\mathcal{O}$ as if Gaussian null coordinates were defined throughout $\mathcal{O}$.

\medskip

Note that there are some freedom left in construction the Gaussian null coordinates as we were free to choose the vector field $k^a$ and the coordinates $(x^3,x^4)$ on ${\mathcal{Z}}$. For instance, by rescaling of the null vector field $k^a$ on ${\mathcal{Z}}$ as $k^a\rightarrow \widetilde k^a= f\, k^a$, where $f$ is a positive function on ${\mathcal{Z}}$ yields a coordinate transformation 
\begin{equation} 
u\rightarrow \widetilde u=\frac1f \, u\,, \hskip.3cm r\rightarrow \widetilde r=f\,r\,,
\end{equation}
while, by making a coordinate transformation $x^A\rightarrow \widetilde x^A=\widetilde x^A(x^3,x^4)$ on ${\mathcal{Z}}$ yields a transformation of $g_{AB}$ as 
\begin{equation} 
g_{AB}\rightarrow \widetilde g_{AB}=g_{EF}\left(\frac{\partial  x^E}{\partial\widetilde x^A}\right)\left(\frac{\partial  x^F}{\partial\widetilde x^B}\right)\,.
\end{equation} 

\subsection{The basic variables}\label{variables}

In deriving the main results of this paper we are going to apply the techniques of the null characteristic initial value problem, in a great extent by adopting Friedrich's method \cite{friedrich,friedrich1} to the selected Einstein-Maxwell systems. In doing so the Newman-Penrose formalism \cite{newman:penrose} will be used.  
 
\subsubsection{The geometry}\label{gaugefix}

The contravariant form of the spacetime metric (\ref{le1})---in a Gaussian null coordinate system $(u,r,x^3,x^4)$---can be given as
\begin{equation} 
g^{\alpha\beta}=\left( 
\begin{array}{ccc} 
0 & 1 & 0 \\ 
1 & g^{rr} & g^{rB} \\ 
0 & g^{Ar} & g^{AB} 
\end{array} 
\right) .  \label{m2} 
\end{equation} 
By choosing some real-valued functions $U $, $X ^A$ and complex-valued functions $\omega ,$ $\xi ^A$ on ${\mathcal{O}}$ such that 
\begin{equation}\label{m3} 
g^{rr}=2(U -\omega \bar\omega),\ \ g^{rA}=X ^A- (\bar\omega\xi 
^A+\omega \bar\xi^A),\ \ g^{AB}=-(\xi ^A \bar\xi^B+\bar\xi^A\xi ^B), 
\end{equation} 
and setting 
\begin{equation} 
\ell^\mu =\delta ^\mu {}_r,\ \ n^\mu =\delta ^\mu {}_u+U \delta ^\mu 
{}_r+X ^A\delta ^\mu {}_A, \ \ m^\mu =\omega \delta ^\mu {}_r+\xi ^A\delta 
^\mu {}_A,  \label{tet} 
\end{equation} 
we obtain a complex null tetrad $\{\ell^a,n^a,m^a,\overline{m}^a\}$ in ${\mathcal{O}}$ \cite{newman:penrose}. In accordance with the vanishing of $g_{uu}$ and $g_{uA}$ on $\mathcal{H}_2$ the functions $U $, $X ^A$, and $\omega$ are required to vanish there, i.e., 
\begin{equation}\label{uxo} 
U=X^A=\omega=0
\end{equation}
on ${\mathcal{H}}_2$. This, in virtue of (\ref{tet}), guaranties that $n^a$ is tangent to the generators of ${\mathcal{H}}_2$, actually $n^a=k^a$ on ${\mathcal{H}}_2$, and also that $m^a$ and $\overline{m}^a$ are everywhere tangent to the cross-sections ${\mathcal{Z}}_u$ of ${\mathcal{H}}_2$. Note, however, that in general (see Proposition\,\ref{ndl} below) the complex null vector $m^a$ and $\overline{m}^a$ are not parallelly propagated along the generators of ${\mathcal{H}}_2$.

\medskip

The Newman-Penrose equations\footnote{To avoid the steady citation of this fundamental work of Newman and Penrose \cite{newman:penrose}  throughout this paper, as in \cite{holographs1}, the equations referred to as $(NP.n.$`a combination of a number $\&$ a lowercase letter'$)$ are always meant to be the original equations listed as $($n.`a combination of a number $\&$ a lowercase letter'$)$ in \cite{newman:penrose}. In order to assist those readers who would like to check the arguments below these equations are also given explicitly in the appendix using the same numbering pattern.}$^{,}$\footnote{It is known that the Newman-Penrose formalism---in particular, the definitions of the basic variables---appear to be more natural when dealing with spinors rather than with tetrads. As the justification of Theorem \ref{equivalent} (see also Appendix A) will require the use of spinorial forms the definitions are recalled in both tetrad and spinorial notations.} involves derivatives of spin coefficients 
(see Table\,\ref{table:data0}),
\begin{table}[h!]
\centering \small \vskip-.15cm
\begin{tabular}{lll} 
%\hline 
$\phantom{\frac{\frac12}{\frac12}}$\hskip-0.1cm $\kappa=\Gamma_{00'00}=\ell^am^b\nabla _a\ell_b$ & $\varepsilon=\Gamma_{00'01}=\frac 12 \ell^a(n^b\nabla _a\ell_b-% 
\overline{m}^b\nabla_am_b)$ & $\pi=\Gamma_{00'11}=-\ell^a\overline{m}^b\nabla _an_b$ \\ %\hline 
$\phantom{\frac{\frac12}{\frac12}}$\hskip-0.1cm $\rho =\Gamma_{10'00}=\overline{m}^am^b\nabla _a\ell_b$ & $\alpha=\Gamma_{10'01}=\frac 12\overline{m}% 
^a(n^b\nabla _a\ell_b-\overline{m}^b\nabla_am_b)$ & $\lambda=\Gamma_{10'11}=-\overline{m}^a% 
\overline{m}^b\nabla _an_b$ \\ %\hline 
$\phantom{\frac{\frac12}{\frac12}}$\hskip-0.1cm $\sigma=\Gamma_{01'00}=m^am^b\nabla _a\ell_b$ & $\beta=\Gamma_{01'01}=\frac 12m^a(n^b\nabla _a\ell_b-\overline{m}% 
^b\nabla _am_b)$ & $\mu=\Gamma_{01'11}=-m^a\overline{m}^b\nabla _an_b$ \\ %\hline 
$\phantom{\frac{\frac12}{\frac12}}$\hskip-0.1cm $\tau=\Gamma_{11'00}=n^am^b\nabla _a\ell_b$ & $\gamma=\Gamma_{11'01}=\frac 12n^a(n^b\nabla _a\ell_b-\overline{m}% 
^b\nabla _am_b)$ & $\nu =\Gamma_{11'11}=-n^a\overline{m}^b\nabla _an_b$ %\\ %\hline 
\end{tabular} 
\vskip-.25cm
\caption{\small The spin coefficients.}\label{table:data0}
\end{table}\vskip-.15cm \hskip-.5cm 
and also those of the Weyl and Ricci spinor components (see Tables\,\ref{table:dataW} and \ref{table:dataR}), in the direction of the frame vectors defined above.
\begin{table}[h!]
\centering \small \vskip-.15cm
\begin{tabular}{l} 
%\hline 
$\phantom{\frac{\frac12}{\frac12}{4}}$ $\Psi _0=\Psi_{0000}=-C_{abcd}\ell^am^b\ell^cm^d $ \\% \hline 
$\phantom{\frac{\frac12}{\frac12}{4}}$ $\Psi _1=\Psi_{0001}=-C_{abcd}\ell^an^b\ell^cm^d $ \\ %\hline 
$\phantom{\frac{\frac12}{\frac12}{4}}$ $\Psi _2=\Psi_{0011}=-\frac 12C_{abcd}(\ell^an^b\ell^cn^d-\ell^an^bm^c\overline{m}^d)$ \\ %\hline 
$\phantom{\frac{\frac12}{\frac12}{4}}$ $\Psi _3=\Psi_{0111}=-C_{abcd}n^a\ell^bn^c\overline{m}^d $ \\ %\hline 
$\phantom{\frac{\frac12}{\frac12}{4}}$ $\Psi _4=\Psi_{1111}=-C_{abcd}n^a\overline{m}^bn^c\overline{m}^d $ %\\ %\hline 
\end{tabular}
\vskip-.25cm\caption{\small The Weyl spinor components.}\label{table:dataW}
\end{table} %\vskip-.15cm
\begin{table}[h!]
\centering \small \vskip-.15cm
\begin{tabular}{ll} 
%\hline 
$\phantom{\frac{\frac12}{\frac12}{4}}$ $\Phi _{00}=\Phi_{000'0'}=-\frac 12R_{ab}\ell^a\ell^b$ & $\Phi_{01}=\Phi_{000'1'}=-\frac 12R_{ab}\ell^am^b$ \\ 
%\hline 
$\phantom{\frac{\frac12}{\frac12}{4}}$ $\Phi _{22}=\Phi_{111'1'}=-\frac 12R_{ab}n^an^b$ &  $\Phi _{02}=\Phi_{001'1'}=-\frac 12R_{ab}m^am^b$  \\ 
%\hline 
$\phantom{\frac{\frac12}{\frac12}{4}}$ $\Phi _{11}=\Phi_{010'1'}=-\frac 14R_{ab}(\ell^an^b+m^a\overline{m}^b)$ & $\Phi_{12}=\Phi_{011'1'}=-\frac 12R_{ab}n^am^b $ \\ 
%\hline 
$\phantom{\frac{\frac12}{\frac12}{4}}$ $\Lambda=\frac 
1{12}R_{ab}(\ell^an^b-m^a\overline{m}^b)$\,, {\rm with} $\Lambda=\overline{\Lambda}$ & $\Phi_{\beta\alpha}=\overline{\Phi}% 
_{\alpha\beta}$\,, {\rm i.e.} $\Phi_{aba'b'}={\Phi}_{(ab)(a'b')}$%\\ 
%\hline 
\end{tabular} 
\vskip-.25cm\caption{\small The Ricci spinor components.}\label{table:dataR}
\end{table}\vskip-.15cm

\medskip

Denote the corresponding operators in ${\mathcal{O}}$ as 
\begin{equation} 
\mathrm{D}=\partial /\partial r,\ \ \Delta =\partial /\partial u+ U \cdot
\partial /\partial r+X^A \cdot\partial /\partial x^A,\ \  \delta 
=\omega \cdot\partial /\partial r+\xi ^A\cdot\partial /\partial x^A. 
\label{dop} 
\end{equation} 

The Newman-Penrose equation can be simplified by assuming that the tetrad $\{\ell^a,n^a,m^a,\overline{m}^a\}$ is parallelly propagated along the null geodesics with tangent
$\ell^a=\left( \partial /\partial r\right) ^a$ in ${\mathcal{O}}$ \cite{newman:penrose}. Then, in virtue of their definitions (see Table\,\ref{table:data0}) for the spin coefficients $\kappa =\pi =\varepsilon =0$, $\rho = \overline{\rho}$, $\tau =\overline{\alpha }+\beta$ hold everywhere in ${\mathcal{O}}$. Moreover, since $n^a$ was chosen such that $n^a=k^a$, and thereby $n^e\nabla_e n^a=0$ hold along the generators of ${\mathcal{H}}_2$ the spin coefficient $\nu =-n^a\overline{m}^b\nabla _an_b$ vanish on ${\mathcal{H}}_2$. Since $u$ is an affine parameter along the generators of ${\mathcal{H}}_2$ and $\gamma=\frac 12n^a(n^b\nabla _a\ell_b-\overline{m}^b\nabla _am_b)$ we also have that $\gamma+\overline\gamma=0$ along these generators. Finally, by performing a rotation of the form $m^a\rightarrow e^{i\phi}m^a$, where  $\phi: {\mathcal{H}}_2\rightarrow\mathbb{R}$ is a suitably chosen real function, the term $\overline{m}^b\nabla _am_b$ and, in turn, the spin coefficient $\gamma$ can be guaranteed to vanish 
along 
the 
generators of ${\mathcal{H}}_2$, and thereby throughout ${\mathcal{H}}_2$.

\medskip

\subsubsection{The electromagnetic field}%\label{nchfr}

The electromagnetic field, $F_{ab}$, can be represented by making use of the Maxwell spinor components (see Table\,\ref{table:dataM})
\begin{table}[h!]
\centering \small \vskip-.25cm
\begin{tabular}{l} 
%\hline 
$ \phantom{\frac{\frac12}{\frac12}{4}} \phi_0 =\phi_{00} = F_{ab}\ell^am^b $ \\ %\hline 
$\phantom{\frac{\frac12}{\frac12}{4}} \phi_1 =\phi_{01} = \frac12\,F_{ab}\,\left(\ell^an^b +  \overline{m}^a m^b \right)$ \\ %\hline 
$ \phantom{\frac{\frac12}{\frac12}{4}} \phi_2 =\phi_{11} = F_{ab}\overline{m}^an^b\,, $ %\\ 
%\hline
\end{tabular}
\vskip-.25cm\caption{\small The Maxwell spinor components.}\label{table:dataM}
\end{table} \vskip-.15cm \hskip-.5cm 
satisfying the Maxwell equations as given by (NP.A1) of \cite{newman:penrose} (see also  (\ref{maxNPeqs}) and (NP.A1.a-d) in Appendices A and B, respectively). 

By (\ref{ein}) and (\ref{max}) the Ricci spinor components $\Phi_{ij}$---defined by (NP.4.3b)  (see also Table\,\ref{table:dataR}) with indices $i,j$ taking the values $0,1,2$---can be given as
\begin{equation}\label{riccimax}
\Phi_{ij}=k\,\phi_i\overline{\phi}_j\,,
\end{equation} 
with\,\footnote{If one would like to use instead of the geometric units the standard ones the factor $-8\pi$ on the right hand side of (\ref{ein}) has to be replaced by $-\frac{8\pi G}{c^4}$ and then the factor $k$ takes the value $\frac{2 G}{c^4}$. } $k=2$.
Note also that by (NP.4.3b) $\Lambda=\frac1{24}R$. As the energy-momentum tensor (\ref{max}) is trace-free we have that in the present case $R=4\widetilde\Lambda$ and, in turn, $\Lambda$ is one sixth of that of the cosmological constant $\widetilde\Lambda$. 

\section{Setup of the characteristic initial value problem}\label{nchfr}
\setcounter{equation}{0}

In proceeding by specifying the characteristic initial value problem applied in this paper recall, first, that the full set of Newman-Penrose equations, $(NP.6.10a-h), (NP.6.11a-r)$ and $(NP.6.12a-h)$, and the Maxwell equations (NP.A1) of \cite{newman:penrose}, when taking them as a coupled set of first order partial differential equations, with  respect to Gaussian null coordinates, $(u,r,x^3,x^4)$ in ${\mathcal{O}}$, for the components of the vector valued variable\,\footnote{If we had a more generic system of matter fields the quantity $\Lambda=\frac1{24}R$ should also be listed in $\mathbb{V}$. As $\Lambda=\frac1{6}\widetilde\Lambda$ is constant it may be left out in the present case.} 
\begin{equation}\label{V} 
\mathbb{V}=(\xi^A,\omega,X^A,U;\rho,\sigma,\tau,\alpha,\beta,\gamma,\lambda,\mu,\nu;\Psi_0,\Psi_1, \Psi_2,\Psi_3,\Psi_4; \phi_0,\phi_1, \phi_2)^T
\end{equation}
can be seen to comprise an overdetermined system simply because we have more equations than unknowns in $\mathbb{V}$. 
This overdetermined feature requires a careful treatment. {The first adequate resolution of this problem was give} by Friedrich in \cite{friedrich} in the vacuum case. The rest of this section (see also Appendix A) is to work out an analogous treatment relevant for the electrovacuum case with allowing non-zero cosmological constant.

\medskip

In accordance with the observations made first by Bondi and Sachs \cite{bondi, sachs}, whenever we are given the pair of smooth null hypersurfaces ${\mathcal{H}}_1$ and ${\mathcal{H}}_2$ intersecting on a two-dimensional spacelike surface ${\mathcal{Z}}$, some of the Einstein's equations will be ``interior equations''---these are the pertinent constraint equations---on ${\mathcal{Z}}$, ${\mathcal{H}}_1$ and ${\mathcal{H}}_2$, respectively. Therefore, we shall only have limited freedom in specifying the initial values for the variables listed in $\mathbb{V}$.

More concretely, it can be seen then that (NP.6.10f-g) and (NP.6.11k-m) are the ``inner equations'' on ${\mathcal{Z}}$ which, provided that $\rho,\sigma,\mu,\lambda,\tau\, \phi_1, \,\xi^A$ are known on ${\mathcal{Z}}$, can be solved algebraically for the rest of the variables listed in $\mathbb{V}$. As soon as all the components of $\mathbb{V}$ are known on ${\mathcal{Z}}$, the initial data $\mathbb{V}_0$ can be determined on ${\mathcal{H}}_1$ and ${\mathcal{H}}_2$ by integrating a sequence of ordinary differential equations---these are given by the respective set of inner equations, $\{$(NP.6.10a-d), (NP.6.11a-i),(NP.6.12a-d), (NP.A1.a-b)$\}$ and $\{$(NP.6.10e), (NP.6.10h), (NP.6.10j), (NP.6.11n-r), (NP.6.12e-h), (NP.A1.c-d)$\}$---along the null geodesic generators a ${\mathcal{H}}_1$ and ${\mathcal{H}}_2$, respectively. These equations---which involve only inner derivatives on ${\mathcal{H}}_1$ and ${\mathcal{H}}_2$, respectively,---can be integrated successively provided that the Weyl 
and Maxwell spinor components $\Psi_0$, $\phi_0$ on ${\mathcal{H}}_1$ and $\Psi_4$, $\phi_2$ on ${\mathcal{H}}_2$ are known. 

Accordingly, in specifying a full initial data set 
\begin{equation}\label{V0} 
\mathbb{V}_0=\{\xi^A,\omega,X^A,U;\rho,\sigma,\tau,\alpha,\beta,\gamma,\lambda,\mu,\nu;\Psi_0,\Psi_1, \Psi_2,\Psi_3,\Psi_4; \phi_0,\phi_1, \phi_2\}|_{{{\mathcal{H}}_1}\cup{{\mathcal{H}}_2}}\,,
\end{equation}
we have to start with a ``reduced initial data set''\,\footnote{Note that choice for the reduced initial data is not unique. Nevertheless, as the argument would go in an analogous way for all the other possible choices hereafter we shall use only the particular one given as (\ref{Vred}).} $\mathbb{V}_0^{red}$ given as
\begin{equation}\label{Vred} 
\mathbb{V}^{red}_{0}=\{\Lambda\in\mathbb{R}\} \cup \{\rho,\sigma,\mu,\lambda,\tau\,; \phi_1;
\,\xi^A \}|_{{\mathcal{Z}}} \cup
\{\Psi_0; \phi_0\}|_{{{\mathcal{H}}_1}} \cup 
\{\Psi_4; \phi_2\}|_{{{\mathcal{H}}_2}} \,,
\end{equation}
i.e.\,besides fixing $\Lambda=\frac16\widetilde\Lambda$, a reduced set of initial data consist of the Weyl and Maxwell spinor components $\Psi_0$, $\phi_0$ on ${\mathcal{H}}_1$ and $\Psi_4$, $\phi_2$ on ${\mathcal{H}}_2$, and of the spin-coefficients $\rho,\sigma,\tau,\mu,\lambda$, the Maxwell spinor component $\phi_1$, and a complex vector field $\xi^A$, such that $g^{AB}=-(\xi^A\overline{\xi}^B+\overline{\xi}^A\xi^B)$ is a negative definite metric, on ${\mathcal{Z}}$. Once $\mathbb{V}_0^{red}$ is known, by making use of the inner (or constraint) equations on ${\mathcal{H}}_1$ and ${\mathcal{H}}_2$ a full initial data set $\mathbb{V}_0$ can be determined\,\footnote{For the class of spacetimes selected in Section \ref{stac}, the determination $\mathbb{V}_0$ will be carried out in details in Section\,\ref{app}.} as indicated above. Then we have 
\begin{lemma}\label{lHF2} 
Assume that $\mathbb{V}$ is a solution to the coupled Newman-Penrose and Maxwell equations. Denote by $\mathbb{V}_0$ the restriction of $\mathbb{V}$ onto ${\mathcal{H}}_1 \cup {\mathcal{H}}_2$, and by $\mathbb{V}_0^{red}$ the reduced initial data deduced from $\mathbb{V}_0$. Then $\mathbb{V}_0^{red}$, along with the inner Newman-Penrose and Maxwell equations, uniquely determines the initial data set $\mathbb{V}_0$ on ${\mathcal{H}}_1 \cup {\mathcal{H}}_2$.
\end{lemma}

In proceeding, note that, by adopting Friedrich's method to electrovacuum spacetimes, by taking aside some of the Newman-Penrose and Maxwell equations and taking linear combinations of some of them the following ``reduced set'' of Einstein-Maxwell equations can be derived 
\renewcommand{\theequation}{EM.\arabic{equation}}
\setcounter{equation}{0}
\begin{eqnarray}  
&& \mathrm{D}\,\xi ^A={\rho }\,\xi ^A+{\sigma }\,\bar\xi^A \\ 
&& \mathrm{D}\,\omega={\rho\, \omega }+\sigma \,\overline{\omega
  }-\tau  \\  
&& \mathrm{D}X^A={\tau }\,\bar\xi^A+{\bar{\tau}\,\xi }^A \\ 
&& \mathrm{D}\,U ={\tau }\,\overline{\omega }+{\overline{\tau}\,\omega 
    -(\gamma +}  
\overline{\gamma }) \\ 
&& \mathrm{D}\,{\rho} = {\rho }^2+{\sigma \,\overline{\sigma}}
  + \Phi_{00} \\
&& \mathrm{D}\,{\sigma } = 2{\rho \,\sigma }+{{\Psi }_0}  \\  
&& \mathrm{D}\,{\tau } = {\tau
\,\rho }+{\overline{\tau}\,\sigma }+{{\Psi }_1} + \Phi_{01}  \\  
&& \mathrm{D}\,{\alpha}
= {\rho \,\alpha }+{\beta \,\overline{\sigma}} + \Phi_{10}  \\  
&& \mathrm{D}\,\beta = {\alpha \,\sigma }+{\rho \,\beta }+{{\Psi }_1} 
\\  
&& \mathrm{D}\,{\gamma} = {\tau \,\alpha }+{\overline{\tau}\,\beta
}+{{\Psi}_2- \Lambda + \Phi_{11}} \\  
&& \mathrm{D}\,{\lambda} = {\rho\, \lambda
}+{\overline{\sigma}\,\mu + \Phi_{20}} \\  
&& \mathrm{D}\,{\mu } = {\rho \,\mu 
}+{\sigma\,\lambda }+{{\Psi }_2 + 2\,{\Lambda}}\\  
&& \mathrm{D}\,{\nu} =
{\overline{\tau}\,\mu}+{\tau\,\lambda}+{{\Psi}_3 + \Phi_{21}} \\  
&&\hskip-1.cm  \Delta\, \Psi_0 -\delta\,(\Psi
_1{+{\Phi}_{01}})+\mathrm{D}\, {{\Phi 
  }_{02}} =(4\,\gamma -\mu )\,\Psi _0-2\,(2\,\tau+\beta
)\,\Psi_1+3\,\sigma\,\Psi_2 
\\ && \hskip0.5cm 
-{\overline{\lambda}}\,{\Phi }_{00}-2\,\beta\,{\Phi }_{01} 
+2\,\sigma\,{\Phi}_{11}+\rho\,{\Phi }_{02}\nonumber\\
&&\hskip-1.cm   
\Delta\, (\Psi _1-{{\Phi }_{01}}) + \mathrm{D}\,(\Psi _1{-{\Phi
  }_{01}})-\delta\,( \Psi _2+2\,{\Lambda }) +\delta\, {\Phi}_{00}
-\overline{\delta }\,\Psi _0 +\overline{ 
\delta }\,{{\Phi }_{02}}= \\
&& \hskip0.5cm +(\nu-4\,\alpha)\,\Psi_0+2\,(\gamma+2\,
\rho-\mu)\,\Psi_1  -3\,\tau \,\Psi _2+\,2\,\sigma\,\Psi_3 \nonumber\\ 
&& \hskip+0.5cm
   +(2\,\tau-\overline{\nu})\,{\Phi
   }_{00}+2\,(\overline{\mu}-\gamma-\rho)\,{\Phi}_{01} -2\,\sigma
   \,{\Phi 
   }_{10} +2\,\tau\, {\Phi }_{11}+(3\,\alpha 
-\overline{\beta })\,{{\Phi }_{02} -2\,\rho\,{\Phi }_{12}} 
\nonumber \\
&&\hskip-1.cm  \Delta\,(\Psi_2+2\,\Lambda)+ \mathrm{D}\,(\Psi
	  {_2+2\,\Lambda )}-\delta\,(\Psi 
_3+{{\Phi}_{21}})-\overline{\delta }\,(\Psi _1+{{\Phi }_{01}} 
)+\Delta \,{{\Phi }_{00}}+\mathrm{D}\,{\Phi }_{22}= \\
&& \hskip+0.5cm -\lambda\, \Psi _0+2\,(\nu-\alpha )\,\Psi
	  _1+3\,(\rho-\mu)\, \Psi _2  
-2\,{\overline{\alpha }}\,\Psi _3+\sigma\, \Psi_4 \nonumber \\ && \hskip+0.5cm 
+(2\,\gamma +2\,\overline{ 
\gamma }-\overline{\mu })\,{{\Phi }_{00}} {-2\,(\alpha
  +}\overline{\tau })\,{{\Phi }_{01}}  
-{2}\,\tau\,{{\Phi }_{10}}+2\,(\rho-\mu)\,{\Phi }_{11} \nonumber\\
&&\hskip+0.5cm -\overline{\lambda }\,{\Phi }_{20}  
+\overline{\sigma }\,{{\Phi }_{02}}+{2\,\beta\,{\Phi
  }_{21}}+\rho\,{\Phi }_{22}\nonumber \\  
&&\hskip-1.cm  \Delta\, (\Psi _3{-{\Phi }_{21}}) + \mathrm{D}\,(\Psi 
_3{-{\Phi }_{21}})-\delta\,\Psi_4-\overline{\delta }\,(\Psi
_2+2\,\Lambda )  +\delta\,{\Phi }_{20} +\overline{\delta }\,{{\Phi } 
_{22}} = \\
&& \hskip+0.5cm
-2\,\lambda\, \Psi _1+3\,\nu\, \Psi _2-2\,(\gamma +2\,\mu -\rho)\,\Psi
_3+(4\,\beta -\tau )\,\Psi _4 \nonumber\\ &&  \hskip+0.5cm
+2\,\mu \,{{\Phi }}_{10}-{(2\,\beta }-2\,\overline{\alpha
}+\nu)\,{{\Phi }_{20}}-2\,\nu \,{{\Phi }_{11}}  
+{2\,\lambda\, {\Phi }_{12}+2\,(\gamma + 
\overline{\mu }-\rho)\,{{\Phi }_{21}-}}\overline{\tau
}\,{{{\Phi}_{22}}}\nonumber \\  
&&\hskip-1.cm \mathrm{D}\,\Psi_4-\overline{\delta}\,(\Psi _3+{{\Phi 
  }_{21}})+\Delta\,{{  
\Phi }_{20}} =
-3\,\lambda\, \Psi _2+2\,\alpha\, \Psi _3+\rho\, \Psi _4 \\
&&\hskip+0.5cm +2\,\nu\, {{\Phi }  
}_{10}-2\,\lambda\, {\Phi }_{11}   
-{(}2\,\gamma -2\,\overline{\gamma }+\overline{\mu })\,{{\Phi }_{20}} 
{-2\,(\overline{\tau }-\alpha )\,{\Phi }_{21}+}\overline{\sigma
}\,{{\Phi }_{22}}  \nonumber  \\
%\end{eqnarray} 
%\begin{eqnarray} 
&& \Delta \phi_0 - \delta \phi_1 = (2\,\gamma - \mu)\,\phi_0 -
2\,\tau\,\phi_1 + \sigma\,\phi_2\\
&& \Delta \phi_1+ \mathrm{D}\phi_1 - \delta \phi_2 - \overline{\delta}
\phi_0=  (\nu- 2\,\alpha)\,\phi_0 + 2\,(\rho - \mu)\,\phi_1 
-(\tau - 2\,\beta)\,\phi_2 \\
&& \mathrm{D} \phi_2 - \overline{\delta} \phi_1 = -\lambda\,\phi_0 +
\rho\,\phi_2\,.  
\end{eqnarray} 
\renewcommand{\theequation}{3.\arabic{equation}}
\setcounter{equation}{7}{}
As it is discussed in more details in Appendix A these equations can be divided into various subclasses. For instance, (EM.1) - (EM.4) are yielded by applying the commutation relations---these are equivalent to the vanishing of some of the components of the torsion tensor that can be associated with the connection $\nabla_a$---to the Gaussian null coordinates $u,r,x^A$, while (EM.5) - (EM.13) stands for some of the Ricci identities. Finally, (EM.14) - (EM.18) and (EM.19) - (EM.21) can be seen to be combinations of some of the Bianchi identities and some of the Maxwell equations, respectively. All of these subsets are more complicated if the gauge choices made in subsection \ref{gaugefix} are not applied. 

\medskip

The following two theorems---the proofs of which are given separately in Appendix \ref{Appendix A}---justify that, from evolutionary point of view, the reduced system (EM.1) - (EM.21) is as good as the complete set of the Newman-Penrose and Maxwell equations. 

\begin{theorem}\label{unique} 
Denote by $\mathbb{V}_0$ a full initial data set, satisfying the ``inner'' Newman-Penrose and Maxwell equations on the initial data surface comprised by the pair of intersecting null hypersurfaces ${\mathcal{H}}_1$ and ${\mathcal{H}}_2$. Then, there exist a unique solution, $\mathbb{V}$, on the domain of dependence $D[{\mathcal{H}}_1\cup {\mathcal{H}}_2]$, to the reduced Einstein-Maxwell equations, (EM.1) - (EM.21), such that $\mathbb{V}\vert_{{{\mathcal{H}}_1\cup {\mathcal{H}}_2}}=\mathbb{V}_0$.
\end{theorem} 

\begin{theorem}\label{equivalent} 
Consider an initial data specification $\mathbb{V}_0$, satisfying the inner Newman-Penrose and Maxwell equations on the initial data surface, and denote by $\mathbb{V}$ the associated unique solution to the reduced set of Einstein-Maxwell equations. Then, $\mathbb{V}$ is also a solution to the full set of the coupled Newman-Penrose and Maxwell equations. 
\end{theorem}

By combining the above results and observations we have the following 

\begin{theorem}\label{HF2} 
To a reduced initial data set $\mathbb{V}_0^{red}$ % [as specified in (\ref{Vred})], in the characteristic initial value problem, 
there always exists (up to diffeomorphisms) a unique solution %, in $D[\mathcal{H}_1\cup\mathcal{H}_2]$, 
to the source-free Einstein-Maxwell equations allowing non-zero cosmological constant.
\end{theorem} 

\section{The determination of a full initial data set $\mathbb{V}_0$}\label{app} 
\renewcommand{\theequation}{4.\arabic{equation}}
\setcounter{equation}{0}

In virtue of Theorem\,\ref{unique} a reduced initial data set uniquely determines the corresponding solution of (EM.1) - (EM.21). As we have seen, the reduced set of initial data $\mathbb{V}_0^{red}$ consist of the Weyl and Maxwell spinor components $\Psi_0$, $\phi_0$ on ${\mathcal{H}}_1$ and $\Psi_4$, $\phi_2$ on ${\mathcal{H}}_2$, and of the spin-coefficients $\rho,\sigma,\tau,\mu,\lambda$, the Maxwell spinor component $\phi_1$, and a complex vector field $\xi^A$, such that $g^{AB}=-(\xi^A\overline{\xi}^B+\overline{\xi}^A\xi^B)$ is a negative definite metric, on ${\mathcal{Z}}$. In addition, the value of $\Lambda$, given in terms of the cosmological constant as $\Lambda=\frac16\widetilde\Lambda$, also have to be specified. 

\medskip

Note that the {distorted black hole} spacetimes---with a pair of null hypersurfaces, $\mathcal{H}_1$ and $\mathcal{H}_2$, generated by expansion and shear free geodesically complete null congruences such that they intersect on a two-dimensional spacelike surface, $\mathcal{Z}=\mathcal{H}_1\cap\mathcal{H}_2$---are not the most generic ones to which the results of the previous section apply. In virtue of the characterization of null congruences, as given, e.g., in subsection 6.1.2 in \cite{exact}, and by the definition of the spin coefficients $\rho,\sigma,\mu,\lambda$ (see Table\,\ref{table:data0}), the expansion and shear free character of the null generators of ${\mathcal{H}}_1$ and ${\mathcal{H}}_2$ can be seen to be equivalent to the vanishing of $\sigma$ and $\rho$ on ${\mathcal{H}}_1$, and also of $\lambda$ and $\mu$ on ${\mathcal{H}}_2$. Taking into account $(NP.6.11j)$, $(NP.6.11m)$ and $(NP.6.11n)$, the vanishing of $\lambda$ and $\mu$ on ${\mathcal{H}
}_2$, along with the vanishing of 
$\gamma$ and $\nu$ 
there, can be seen to be equivalent to the vanishing of $\Psi_4$, $-\Psi_3+\Phi _{21}$ and $\Phi _{22}$ on ${\mathcal{H}}_2$, and similarly, by $(NP.6.11a)$, $(NP.6.11b)$, $(NP.6.11k)$ the vanishing of $\sigma$ and $\rho$ on ${\mathcal{H}}_1$ imply the 
vanishing of $\Psi_0$, $-\Psi_1+\Phi _{01}$ and $\Phi_{00}$ on ${\mathcal{H}}_1$, respectively. Applying then the relations $\Phi _{ij}=k\,\phi_{i} \overline{\phi}_{j}$, with $i,j=0,1,2$, the expansion and shear free character of the null generators of ${\mathcal{H}}_1$ and ${\mathcal{H}}_2$ can be seen to be equivalent to the vanishing of $\Psi_4$, $\Psi_3$ and $\phi_{2}$ on ${\mathcal{H}}_2$, and $\Psi_0$, $\Psi_1$ and $\phi_{0}$ on ${\mathcal{H}}_1$, respectively. 

\medskip

According to these observations the reduced  initial data set, $\mathbb{V}_0^{red}$ given as (\ref{Vred}), simplifies considerably. In particular, the only not identically zero functions which can yet be freely specified are the $\tau$ spin coefficient, the Maxwell spinor component $\phi_1$, and a suitable complex vector field $\xi^A$, all restricted onto ${\mathcal{Z}}$.

\medskip

Note that in determining the full initial data set $\mathbb{V}_0$ we could slightly shorten the argument below by using the fact that ${\mathcal{H}}_1$ and ${\mathcal{H}}_2$ are comprised by expansion and shear free null geodesic congruences. Nevertheless, in order to indicate how one would proceed in the ``generic case'' we shall carry out the determination of a full initial data set $\mathbb{V}_0$, by starting with the reduced one
\begin{equation} 
\mathbb{V}^{red}_0=\{\not\hskip-.1cm\rho,\not\hskip-.1cm\sigma,
\not\hskip-.1cm\mu,\not\hskip-.1cm\lambda,\tau\,;\,\phi_1; 
\,\xi^A\}|_{{\mathcal{Z}}} \cup \{\not\hskip-.1cm\Psi_0;\not\hskip-.1cm\phi_0\}|_{{{\mathcal{H}}_1}} \cup 
\{\not\hskip-.1cm\Psi_4;\not\hskip-.1cm\phi_2\}|_{{{\mathcal{H}}_2}}
\,,
\end{equation} 
where all the crossed quantities are identically zero on the indicated subsets of $\mathcal{H}_1\cup\mathcal{H}_2$.

\medskip

Start by inspecting the ``\,inner equations\,'' on ${\mathcal{Z}}$. Notice first that by $(NP.6.11k)$ and $(NP.6.11m)$, and by the vanishing of $\rho,\sigma,\mu,\lambda$ on ${\mathcal{Z}}$, both $\Psi_1$ and $\Psi_3$ vanish on ${\mathcal{Z}}$.  Furthermore, $(NP.6.10f)$, along with our gauge conditions $\tau=\overline{\alpha}+\beta$ in ${\mathcal{O}}$, gives that 
\begin{equation}\label{beta}
\delta \overline{\xi}^A - \overline{\delta} \xi^A= \left[2\,\overline{\beta}-\overline{\tau}\right]\,\xi^A +\left[\tau - 2\,\beta\right]\,\overline{\xi}^A\,.  
\end{equation}
This equation can be solved algebraically for $\beta$ and, whence also, for $\alpha=\overline{\tau}-\overline{\beta}$ on ${\mathcal{Z}}$. By applying
then $(NP.6.11l)$ we immediately get
\begin{equation} 
\Psi_2=-\delta{\alpha} + \overline{\delta} \beta + \alpha\,\overline{\alpha}
-2\, \alpha\, \beta + \beta\,\overline{\beta} + \Lambda + k\,\phi_1\,\overline{\phi}_1\, 
\label{psi2}  
\end{equation}
which fixes the value of $\Psi_2$ on ${\mathcal{Z}}$, and, in turn, we also have the full initial data on ${\mathcal{Z}}$ as 
\begin{equation}\label{V02} 
\mathbb{V}_0=\{\xi^A,\not\hskip-.1cm\omega,\not\hskip-.1cm X^A,\not\hskip-.1cmU;\not\hskip-.1cm\rho,\not\hskip-.1cm\sigma,\tau,\alpha,\beta,\not\hskip-.1cm\gamma,\not\hskip-.1cm\lambda,\not\hskip-.1cm\mu,\not\hskip-.1cm\nu;\not\hskip-.1cm\Psi_0,\not\hskip-.1cm\Psi_1, \Psi_2,\not\hskip-.1cm\Psi_3,\not\hskip-.1cm\Psi_4; \not\hskip-.1cm\phi_0,\phi_1, \not\hskip-.1cm\phi_2\}|_{\mathcal{Z}}\,.
\end{equation}

\medskip

Consider now the inner equations on ${{\mathcal{H}}_1}$. First of all, since $\Psi_0\equiv 0$ and $\phi_0\equiv 0$ on ${{\mathcal{H}}_1}$ $(NP.6.12a)$, along with the fact that $\Psi_1|_{{\mathcal{Z}}}\equiv 0$, implies that $\Psi_1\equiv 0$ on ${{\mathcal{H}}_1}$. Similarly,  since
$\rho|_{{\mathcal{Z}}}\equiv 0$ and $\sigma|_{{\mathcal{Z}}}\equiv 0$, $(NP.6.11a)$ and $(NP.6.11b)$
imply that $\rho\equiv 0$ and $\sigma\equiv 0$ on ${{\mathcal{H}}_1}$. The vanishing of $\rho$, $\sigma$, $\Phi_{01}$, $\Phi_{10}$ and
$\Psi_1$ on ${{\mathcal{H}}_1}$ can then be used, along with $(NP.610a)$, $(NP.611c-d-e)$, to conclude that 
\begin{equation} 
D\xi^A=\mathrm{D}\alpha=\mathrm{D}\beta=\mathrm{D}\tau=0
\label{dD} 
\end{equation}
and by $(NP.610b-c)$
\begin{eqnarray}
\omega&&\hskip-.6cm=-r\,\tau\\
X^A&&\hskip-.6cm=r\,\left[\tau\,\overline{\xi}^A+\overline{\tau}\,\xi^A\right]
\end{eqnarray}
on ${{\mathcal{H}}_1}$. Similarly, $(NP.6.12b)$, along with the vanishing of $\Delta(\Phi_{00})=k\,[(\Delta{\phi}_0)\overline\phi_0\, +\phi_0\,(\Delta\overline{\phi}_0)]$ on ${{\mathcal{H}}_1}$, gives that
\begin{equation}\label{psi2n2}
\mathrm{D}\Psi_2=0
\end{equation} 
on ${{\mathcal{H}}_1}$. By $(NP.6.11g)$ and by the vanishing of $\phi_0$ we also have then that $\lambda\equiv 0$ on ${{\mathcal{H}}_1}$ since $\lambda$ vanishes on ${\mathcal{Z}}$. Two of other spin coefficients, $\gamma$ and $\mu$, can be determined with the help of $(NP.6.11f)$ and $(NP.6.11h)$ which, along with (\ref{psi2n2}) and their vanishing on ${\mathcal{Z}}$, yield that on ${{\mathcal{H}}_1}$ 
\begin{equation}\label{mu}
\mu=r\,\left[\Psi_2 + 2\,\Lambda\right]\,,
\end{equation}
and
\begin{equation}\label{gamma}
\gamma=r\, \left[\tau\,\alpha+ \overline{\tau}\,\beta +\Psi_2 - \Lambda+ k\,\phi_1\,\overline{\phi}_1\, \right]\,. 
\end{equation}

Finally, as both $\rho$ and $\sigma$ identically vanish on ${{\mathcal{H}}_1}$, and also (\ref{uxo}) holds on ${\mathcal{Z}}$, in virtue of $(NP.6.10d)$ and (\ref{gamma}) we have that the relations 
\begin{equation}
U=-r^2\,\left[2\,\tau\,\overline{\tau}+\frac12\left(\Psi_2+\overline{\Psi}_2\right)-\Lambda+k\,\phi_1\,\overline{\phi}_1\right]
\end{equation}   
hold on ${{\mathcal{H}}_1}$.

\bigskip

By a completely analogous argument, the inner equations on ${{\mathcal{H}}_2}$, along with our gauge choice guaranteeing the vanishing of $\nu$ and $\gamma$ there, can be used to justify the followings. First, since $\Psi_4\equiv 0$ there $(NP.6.12h)$, along with the fact that $\phi_2|_{{\mathcal{H}}_2}\equiv 0$ and 
$\Psi_3|_{{\mathcal{Z}}}\equiv 0$, implies that $\Psi_3\equiv 0$ on ${{\mathcal{H}}_2}$. Similarly,  since $\mu|_{{\mathcal{Z}}}\equiv 0$ and $\lambda|_{{\mathcal{Z}}}\equiv 0$, $(NP.6.11n)$ and $(NP.6.11j)$ imply that $\mu\equiv 0$ and $\lambda\equiv 0$ on ${{\mathcal{H}}_2}$. The vanishing of $\mu$, $\lambda$, $\gamma$, $\nu$ and $\Psi_3$ on ${{\mathcal{H}}_2}$, along with (\ref{uxo}), $(NP.6.10e)$ and $(NP.611r-o-p)$,  can be used then to conclude that
\begin{equation} 
\Delta \xi^A=\Delta \alpha=\Delta \beta=\Delta \tau =0
\label{dDen1} 
\end{equation}
on ${{\mathcal{H}}_2}$. Similarly, $(NP.6.12g)$, along with the vanishing of $\lambda,\mu$ and $\phi_2$ on ${{\mathcal{H}}_2}$, gives then 
\begin{equation}\label{psi2n1}
\Delta\Psi_2=0
\end{equation} 
on ${{\mathcal{H}}_2}$. 
In virtue of $(NP.6.11p)$ we also have that $\Delta\sigma=\delta\,\tau - 2\,\beta\,\tau$ on ${{\mathcal{H}}_2}$ which, along with the vanishing of $\sigma$ on ${{\mathcal{Z}}}$, implies that
\begin{equation}\label{sigma} 
\sigma=u\, \left[\delta\,\tau - 2\,\beta\,\tau\right]%\,.
\end{equation}
holds on ${{\mathcal{H}}_2}$. The only remaining non-trivial spin coefficient $\rho$ gets to be determined on ${{\mathcal{H}}_2}$ by $(NP.6.11q)$,  which, along with (\ref{dDen1}), (\ref{psi2n1}) and the vanishing of $\rho$ on ${\mathcal{Z}}$, implies that 
\begin{equation}\label{rhon2} 
\rho= u\, \left[\,\overline{\delta}\,\tau - 2\,\alpha\,\tau - \Psi_2-2\,\Lambda\right]\, 
\end{equation}
on ${{\mathcal{H}}_2}$.

\bigskip

To have a full initial data set $\mathbb{V}_0$ on ${{\mathcal{H}}_1} \cup {{\mathcal{H}}_2}$, in addition to what we already have the Weyl spinor components ${\Psi_1}$, ${\Psi_0}$ on ${{\mathcal{H}}_1} \cup {{\mathcal{H}}_2}$, along with the Maxwell spinor components $\phi_2$ on ${{\mathcal{H}}_1}$ and $\phi_0$ on ${{\mathcal{H}}_2}$, and also, $\nu$ on ${{\mathcal{H}}_1}$,  are to be known.

In determining ${\phi_0}$ on ${{\mathcal{H}}_2}$ note first that by (NP.A1.d) we have that 
\begin{equation}
\Delta \phi_1|_{{\mathcal{H}}_2}\equiv 0
\end{equation}
Similarly, by (NP.A1.c) we have that $\Delta\phi_0=\delta\,\phi_1 - 2\,\tau\,\phi_1$ on ${{\mathcal{H}}_2}$ which, along with ${\phi_0}|_{{\mathcal{Z}}}=0$ yields then that 
\begin{equation}\label{phi0n1}
\phi_0=u\,\left[\delta\,\phi_1 - 2\,\tau\,\phi_1\right]\,.
\end{equation} 
Now, by making use of $(NP.6.12f)$ we get that $\Delta \Psi_1 - k\,(\Delta\phi_0)\,\overline{\phi}_1 - \delta\,\Psi_2 = - 3\,\tau\,\Psi_2+2\,k\,\tau\,\phi_1\,\overline{\phi}_1$ on ${{\mathcal{H}}_2}$, which, along with (\ref{phi0n1}), ${\Psi_1}|_{{\mathcal{Z}}}=0$ and the $u$-independents of $\phi_1,\tau$ and $\Psi_2$, implies that
\begin{equation}\label{Psi1n1}
\Psi_1=u\,\left[\delta\Psi_2 -3\,\tau \,\Psi_2+k\,(\delta\,\phi_1)\,\overline{\phi}_1  \right]
\end{equation} 
on ${{\mathcal{H}}_2}$. 

Before applying an analogous argument, based on the use of $(NP.6.12e)$ to determine $\Psi_0$ on ${{\mathcal{H}}_2}$, note that by (NP.A1.b) we have that 
\begin{equation}\label{DOphi2}
\mathrm{D}\overline{\phi}_2=\delta\,\overline{\phi}_1
\end{equation} 
on ${{\mathcal{H}}_2}$. This, along with $(NP.6.12e)$, ${\Psi_0}|_{{\mathcal{Z}}}=0$ and the $u$-independents of the coefficients of various terms in $(NP.6.12e)$ on ${{\mathcal{H}}_2}$, implies that 
\begin{equation}\label{Psi0n1}
\Psi_0=\frac{1}{2}\,u^2\left[\delta^2\Psi_2 - (7\,\tau+2\,\beta)\,\delta\Psi_2 + 12\,\tau^2 \Psi_2 + 2\,k\,\left[\delta^2\phi_1-\left(3\,\tau+2\,\beta\right)\delta\phi_1  \right]\,\overline{\phi}_1+ k\,(\delta{\phi_1})\,\delta\overline{\phi}_1\, \right]
\end{equation} 
holds on ${{\mathcal{H}}_2}$.

Completely parallel to the reasoning applied above, the complex conjugate of (\ref{DOphi2}), along with ${\phi_2}|_{{\mathcal{Z}}}=0$, implies that 
\begin{equation}\label{phi2n2}
\phi_2|_{{\mathcal{H}}_1}=r\,\overline{\delta}\,\phi_1 \,.
\end{equation} 
Similarly, in virtue of (NP.A1.a) and the vanishing of ${\phi_0}$ and $\rho$ on ${{\mathcal{H}}_1}$ we also get that 
\begin{equation}\label{Dphi1n1}
\mathrm{D}{\phi_1}|_{{\mathcal{H}}_1}=0\,.
\end{equation}  
By applying  (\ref{DOphi2}) and (\ref{Dphi1n1}), along with the $r$-independence of $\alpha,\beta,\tau$ and ${\phi_1}$  on ${{\mathcal{H}}_1}$ and ${\Psi_3}|_{{\mathcal{Z}}}=0$, we get by $(NP.6.12c)$ that  
\begin{equation}\label{psi3n2}
\Psi_3=r\,\left[\,\overline{\delta}\Psi_2 + k\,(\overline{\delta}{\phi_1})\,\overline{\phi}_1\, \right]
\end{equation} 
on ${{\mathcal{H}}_1}$. 

Similarly, in virtue of (NP.A1.c) and the vanishing of ${\phi_0}$ and $\sigma$ on ${{\mathcal{H}}_1}$ we have that 
\begin{equation}
\Delta{\overline{\phi}_0}|_{{\mathcal{H}}_1}=\overline{\delta}\,\overline{\phi}_1 - 2 \overline{\tau}\,\overline{\phi}_1\,.
\end{equation}  
This, along with $(NP.6.12d)$, ${\Psi_4}|_{{\mathcal{Z}}}=0$ and the $r$-independence of $\alpha,\beta,\tau$ and ${\phi_1}$ on ${{\mathcal{H}}_1}$, implies 
that 
\begin{equation}\label{psi4n2}
\Psi_4=\frac{1}{2}\,r^2\left[\,\overline{\delta}^2\Psi_2 +2\,\alpha\,\overline{\delta}\Psi_2 + 2\,k\,\left(\overline{\delta}^2\phi_1+2\,\alpha\,\overline{\delta}\phi_1  \right)\overline{\phi}_1+ k\,(\overline{\delta}{\phi_1})\,\overline{\delta}\,\overline{\phi}_1 \right]   
\end{equation}
hold on ${{\mathcal{H}}_1}$. 

Finally, by making use of $(NP.6.11i)$, (\ref{mu}), (\ref{phi2n2}) and (\ref{psi3n2}) we have that
\begin{equation}
D\nu|_{{\mathcal{H}}_1}=\,r\,\left[\,\overline{\tau}\,\left(\,\Psi_2 +2\Lambda \,\right)+\left(\,\overline{\delta}\Psi_2 + k\,(\overline{\delta}{\phi_1})\,\overline{\phi}_1\right) + k\,(\overline{\delta}{\phi_1})\,\overline{\phi}_1 \,\right]\,,
\end{equation}
which in turn gives that on  ${{\mathcal{H}}_1}$ %(\overline{\delta}{\phi_1})=\phi_2 !!!!!
\begin{equation}\label{nun1}
\nu=\,\frac{1}{2}\,r^2\left[\,\overline{\delta}\Psi_2  + \overline{\tau}\,\left(\Psi_2 +2\Lambda \right)+ 2\,k\,(\overline{\delta}{\phi_1})\,\overline{\phi}_1\,\right]\,.
\end{equation}

\bigskip

All the above derived relations are collected in Table\,\ref{table:data}\,, 
\begin{table}[h!]
\centering \small \hskip-.15cm
\begin{tabular}{|l|l|l|} 
\hline ${{\mathcal{H}}_1}$ &  ${{\mathcal{Z}}}$
$\phantom{\frac{\frac12}{A}}$ & ${{\mathcal{H}}_2}$ \\ \hline \hline
 $\mathrm{D}\xi^A =0$ &  $\xi^A$ $\phantom{\frac{\frac12}{A}}$ \hfill (data)  & $\Delta\xi^A =0$ \\  \hline 
 $\omega =-r\,\tau$ &  $\omega =0$  $\phantom{\frac{\frac12}{A}}$ \hfill $\leftarrow$  & $\omega = 0$  \hfill (geometry) \\  \hline 
 $X^A=r\,[\tau\,\overline{\xi}^A+\overline{\tau}\,\xi^A]$ $\phantom{\frac{\frac12}{A}}$ &  $X^A=0$ $\phantom{\frac{\frac12}{A}}$ \hfill $\leftarrow$  & $X^A=0$ \hfill (geometry) \\  \hline 
 $U = -r^2\,\widetilde U$ &  $U =0$ $\phantom{\frac{\frac12}{A}}$ \hfill $\leftarrow$  &  $U = 0$ \hfill (geometry) \\  \hline
 $\rho =0$ &  $\rho =0$ $\phantom{\frac{\frac12}{A}}$ & $\rho = u\, \left[\,\overline{\delta}\tau
 - 2\,\alpha\,\tau - \Psi_2-2\,\Lambda\,\right]$ \\  \hline 
 $\sigma=0$  &
 $\sigma=0$ $\phantom{\frac{\frac12}{A}}$ &
 $\sigma=u \, \left[\,\delta\tau - 2\,\beta\,\tau\,\right]$ \\ \hline 
 $\mathrm{D}\tau =0$ &  $\tau$ $\phantom{\frac{\frac12}{A}}$ \hfill (data)  & $\Delta\tau =0$ \\  \hline 
 $\mathrm{D}\alpha=\mathrm{D}\beta=0$ & $\alpha,\beta, \tau=\overline{\alpha}+\beta$ $\phantom{\frac{\frac12}{A}}$ &
  $\Delta \alpha=\Delta \beta=0$ \\ \hline
  $\gamma=r \, \widetilde{\gamma}$ & $\gamma=0$ $\phantom{\frac{\frac12}{A}}$ \hfill $\leftarrow$ & $\gamma=0$  \hfill (gauge) \\ \hline 
 $\mu = r \, \left[\,\Psi_2 +2 \Lambda\,\right]$ & $\mu =0$ $\phantom{\frac{\frac12}{A}}$ & $\mu = 0$  \\ \hline
 $\lambda=0$ & $\lambda=0$ $\phantom{\frac{\frac12}{A}}$ & $\lambda=0$ \\ \hline
 $\nu=\frac{1}{2}\,r^2\,\widetilde{\nu}$ &  $\nu=0$ $\phantom{\frac{\frac12}{A}}$ \hfill $\leftarrow$  & 
 $\nu=0$ \hfill (gauge) \\ \hline
 $\Psi_0=0$ & $\Psi_0=0$ $\phantom{\frac{\frac12}{A}}$ & $\Psi_0=\frac{1}{2}\,u^2\,\widetilde{\Psi}_0$ \\ \hline
  $\Psi_1=0$ &
 $\Psi_1=0$ $\phantom{\frac{\frac12}{A}}$ & $\Psi_1=u\,\widetilde{\Psi}_1$ \\ \hline
 $\mathrm{D} \Psi_2 =0$ &
 $\xi^A, \tau,\phi_1,\Lambda$ \hskip-.4cm $\phantom{\frac{\frac12}{A}}$
 $\rightarrow \ \alpha,\beta,\Psi_2$ & $\Delta\Psi_2=0$ \\ \hline
 $\Psi_3=r\,\widetilde{\Psi}_3 $& $\Psi_3=0$ $\phantom{\frac{\frac12}{A}}$ & $\Psi_3=0$ 
  \\ \hline 
 $\Psi_4=\frac{1}{2}\,r^2\,\widetilde{\Psi}_4$& $\Psi_4=0$ $\phantom{\frac{\frac12}{A}}$ & $\Psi_4=0$ 
  \\ \hline
  $\phi_0 =0$  & $\phi_0 =0$ $\phantom{\frac{\frac12}{A}}$ & $\phi_0=u\,\left[\,\delta\phi_1 - 2\,\tau\,\phi_1\,\right]$ \\ \hline 
 $\mathrm{D}\phi_1 =0$  & $\phi_1$ $\phantom{\frac{\frac12}{A}}$ \hfill (data)  & $\Delta \phi_1 =0$
 \\ \hline 
 $\phi_2 =r\,\overline{\delta}\phi_1$  & $\phi_2=0$ $\phantom{\frac{\frac12}{A}}$ & $\phi_2 =0$
 \\ \hline 
\end{tabular}
\caption{\small The full initial data set ${\mathbb{V}}_{0}$ on the intersecting null hypersurfaces $\mathcal{H}_1\cup\mathcal{H}_2$. }\label{table:data}
\end{table}
where
\begin{eqnarray}
&&\hskip-1cm
\widetilde{\Psi}_0=\delta^2\Psi_2 - (7\,\tau+2\,\beta)\,\delta\Psi_2 + 12\,\tau^2 \Psi_2 + 2\,k\,\left[\delta^2\phi_1-\left(3\,\tau+2\,\beta\right)\delta\phi_1  \right]\,\overline{\phi}_1+ k\,(\delta{\phi_1})\,\delta\overline{\phi}_1  \\
&&\hskip-1cm
\widetilde{\Psi}_1=\delta\Psi_2 -3\,\tau \,\Psi_2+k\,(\delta\,\phi_1)\,\overline{\phi}_1 \\ 
&&\hskip-1cm
\Psi_2=-\delta{\alpha} + \overline{\delta} \beta + \alpha\,\overline{\alpha}
-2\, \alpha\, \beta + \beta\,\overline{\beta} + \Lambda + k\,\phi_1\,\overline{\phi}_1 \\
&&\hskip-1cm
\widetilde{\Psi}_3=\overline{\delta}\Psi_2 + k\,(\overline{\delta}{\phi_1})\,\overline{\phi}_1 \label{Psi3}\\ 
&&\hskip-1cm
\widetilde{\Psi}_4=\overline{\delta}^2\Psi_2 +2\,\alpha\,\overline{\delta}\Psi_2 + 2\,k\,\left(\overline{\delta}^2\phi_1+2\,\alpha\,\overline{\delta}\phi_1  \right)\overline{\phi}_1+ k\,(\overline{\delta}{\phi_1})\,\overline{\delta}\,\overline{\phi}_1 \label{Psi4}\\
&&\hskip-1cm\hskip.23cm
\widetilde{U}=2\,\tau\,\overline{\tau}+\frac12\left(\Psi_2+\overline{\Psi}_2\right)-\Lambda+k\,\phi_1\,\overline{\phi}_1\\
&&\hskip-1cm
\hskip.23cm\widetilde{\gamma}=\tau\,\alpha+ \overline{\tau}\,\beta +\Psi_2 - \Lambda+ k\,\phi_1\,\overline{\phi}_1\\ 
&&\hskip-1cm
\hskip.23cm\widetilde{\nu}=\overline{\delta}\Psi_2 + \overline{\tau}\,\left[\,\Psi_2 +2\,\Lambda \,\right] + 2\,k\,(\overline{\delta}{\phi_1})\,\overline{\phi}_1\,.
\end{eqnarray} 
Notice that by now the explicit the functional form of the variables listed in ${\mathbb{V}}$ are known, on the initial data surface, for all the selected type of spacetimes.

In virtue of all above we get

\begin{theorem}\label{IRI}
Suppose that $(M,g_{ab})$ is an electrovacuum spacetime with Maxwell field $F_{ab}$, allowing a non-zero cosmological constant and possessing a pair of null hypersurfaces $\mathcal{H}_1$ and $\mathcal{H}_2$ generated by expansion and shear free geodesically complete null congruences such that they intersect on a two-dimensional spacelike surface, $\mathcal{Z}=\mathcal{H}_1\cap\mathcal{H}_2$. Then, both the metric $g_{ab}$ and the Maxwell field $F_{ab}$ get to be uniquely determined (up to diffeomorphisms) on the domain of dependence $D[{\mathcal{H}}_1\cup {\mathcal{H}}_2]$ of $\mathcal{H}_1$ and $\mathcal{H}_2$, once a complex vector field $\xi^A$ [determining the induced metric on $\mathcal{Z}$ as $g^{AB}=-(\xi^A\overline{\xi}^B+\overline{\xi}^A\xi^B)$], the $\tau$ spin coefficient and the $\phi_1$ Maxwell spinor component  are specified on $\mathcal{Z}$.
\end{theorem} 

\section{Parallelly propagated curvature blow up}\label{sing}
\renewcommand{\theequation}{5.\arabic{equation}}
\setcounter{equation}{0}

This section is to show that, in general, {\it parallelly propagated} curvature blow up occur along the generators of $\mathcal{H}_1$ and $\mathcal{H}_2$.  To get some insight it is rewarding to inspect Table\,\ref{table:data} for a short while. What might not be salient for the first glance is the $r$-dependence of the Weyl spinor components $\Psi_3$ and $\Psi_4$ along the null generators of ${{\mathcal{H}}_1}$, and similarly, the $u$-dependence of $\Psi_0$ and $\Psi_1$ along the null generators of ${{\mathcal{H}}_2}$. Obviously, all of these quantities vanish at the bifurcation surface but when the asymptotic ends are approached, along the generators of ${{\mathcal{H}}_1}$ or ${{\mathcal{H}}_2}$, both to the future and to the past they may blow up.

\begin{proposition}\label{ndl}
There is a {\it parallelly propagated}\, curvature blow up of the Weyl or Ricci tensor, respectively along the null generators of ${{\mathcal{H}}_1}$ or ${{\mathcal{H}}_2}$, both to the future and to the past, if either of the quantities 
\begin{eqnarray}
&&%&\hskip-.1cm
\widetilde{\Psi}'_0=\delta^2\Psi_2 + (\tau-2\,\beta)\,\delta\Psi_2 + 2\,k\left[\delta^2\phi_1-\left(\tau+2\,\beta\,\right)\delta\phi_1\right]\overline{\phi}_1+ k\,(\delta{\phi_1})\,\delta\overline{\phi}_1 \label{ndl1}\\
&&%&\hskip-.1cm
\widetilde{\Psi}'_1=\delta\Psi_2 +k\,(\delta\,\phi_1)\,\overline{\phi}_1 \label{ndl2}
\end{eqnarray}
$\widetilde{\Psi}_3$, $\widetilde{\Psi}_4$---as given in (\ref{Psi3}) and (\ref{Psi4})---, $\widetilde{\phi}'_0=\delta\phi_1$ or  $\widetilde{\phi}_2=\overline{\delta}\phi_1$ 
is not identically zero on $\mathcal{Z}$.
{The blow up rate is either linear or quadratic.}\,\footnote{{Note that as the generators of ${{\mathcal{H}}_1}$ and ${{\mathcal{H}}_2}$ are maximal null geodesics the maximal blow up rate is known to be quadratic (see, e.g., \cite{kr}).}}  {In particular, unless they vanish identically along the null generators of ${{\mathcal{H}}_1}$ and ${{\mathcal{H}}_2}$, the components $\Psi'_1$, $\Psi_3$, $\Phi'_{01}$ and $\Phi_{12}$ are linear, while $\Psi'_0$, $\Psi_4$, $\Phi'_{00}$ and $\Phi_{22}$ are quadratic functions of the pertinent affine parameters.}
\end{proposition} 
\noindent\textbf{Proof:}{\ } The validity of the above assertions can be justified by inspection of the components of the Weyl and Ricci tensor when they are given with respect to basis fields which are parallelly propagated along the generators of ${{\mathcal{H}}_1}$ and ${{\mathcal{H}}_2}$, respectively. 

%\footnote{
In context of spacetime singularities (note that when these occur they are expected to be in finite affine distance from inner spacetime points) usually orthonormal or pseudo-orthonormal basis fields are applied in qualify a curvature blow up to be a parallelly propagated one \cite{schmidt}. It is straightforward to see that whenever we have a parallelly propagated curvature blow up with respect to a complex null tetrad $\{\ell^a,n^a,m^a,\overline{m}^a \}$ then it is also a  parallelly propagated curvature blow up with respect to the pseudo-orthonormal basis field $\{\ell^a,n^a,Y_{3}^a,Y_{4}^a \}$, where the real spatial unit vectors $Y_{3}^a$ and $Y_{4}^a$ are given as $Y_{3}^a=\frac{1}{\sqrt{2}}(m^a+\overline{m}^a)$ and $Y_{4}^a=\frac{i}{\sqrt{2}}(m^a-\overline{m}^a)$.%}

\medskip

Note first that the complex null tetrad $\{\ell^a,n^a,m^a,\overline{m}^a \}$, due to the gauge fixing applied in Subsection \ref{gaugefix}, is parallelly propagated along the null generators of ${{\mathcal{H}}_1}$. Thereby the blow up of the either of the contractions $\Psi _3=-C_{abcd}n^a\ell^bn^c\overline{m}^d $, $\Psi _4=-C_{abcd}n^a\overline{m}^bn^c\overline{m}^d $ and $\Phi_{22}=-\frac12\,R_{ab} n^an^b=2\,\phi_2\,\overline{\phi}_2$---which occurs whenever either of the quantities $\widetilde{\Psi}_3$, $\widetilde{\Psi}_4$---as given in (\ref{Psi3}) and (\ref{Psi4})---or $\widetilde{\phi}_2=\overline{\delta}\phi_1$ is not identically zero on $\mathcal{Z}$---signifies a true parallelly propagated curvature blow up of the Weyl or Ricci tensor to both ends along the pertinent null generators of ${{\mathcal{H}}_1}$.

\medskip

Consider now the hypersurface ${{\mathcal{H}}_2}$. Note first that in general the vector fields $\ell^a, m^a$ and $\overline{m}^a$ are not parallelly propagated along the null generators of ${{\mathcal{H}}_2}$. To see this note that 
\begin{equation}\label{mder}
n^e\nabla_e m_b={\delta_b}^f \left(n^e\nabla_e m_f\right) = \ell_b\left(n^f n^e\nabla_e m_f\right) + n_b\left(\ell^f n^e\nabla_e m_f\right)-m_b\left(\overline{m}^f n^e\nabla_e m_f\right) - 
\overline{m}_b\left(m^f n^e\nabla_e m_f\right)\,,
\end{equation} 
where %, in virtue of (\ref{tetmet}), 
the relation
%\begin{equation}
${\delta_a}^b=\ell_a n^b+n_a \ell^b-m_a\overline{m}^b-\overline{m}_a m^b$
%\end{equation} 
was applied. The contraction $n^f n^e\nabla_e m^f$ in the first term on the right hand side of (\ref{mder}) vanishes on ${{\mathcal{H}}_2}$ as $n^f n^e\nabla_e m_f=n^e\nabla_e\left( n^f m_f\right)-m_f\left( n^e\nabla_e n^f\right)$ and we also have that $n^f m_f\equiv 0$ and $n^a$ is parallelly propagated along the null generators of ${{\mathcal{H}}_2}$. In the second term $\ell^f n^e\nabla_e m_f$, in virtue of Table\,\ref{table:data0}, reads as $ n^e\nabla_e \left(\ell^f m_f\right)- m_f \left(n^e\nabla_e \ell^f \right)=-\tau$ as $\nabla_e \left(\ell^f m_f\right)$ is identically zero. The third term on the right hand side of (\ref{mder}) vanish on ${{\mathcal{H}}_2}$ as our gauge choice guaranty the term $\overline{m}^f\nabla_e m_f$ to be identically zero along the null generators of ${{\mathcal{H}}_2}$. Finally, the fourth term vanishes as $m^f n^e\nabla_e m_f=\frac12 n^e\nabla_e\left( m^f m_f \right)=0$. In virtue of these observations we get that on ${{\mathcal{H}}_2}$ 
\begin{equation}
n^e\nabla_e m_b=-\tau n_b\,,
\end{equation}
which, along with the fact that $\tau$ is $u$-independent and $n^a$ is parallelly propagated along the null generators of ${{\mathcal{H}}_2}$, implies that the vector field 
\begin{equation}\label{mvesszo}
m'{}^a=m^a+u\, \tau\, n^a\,
\end{equation}
is also parallelly propagated along the null generators of ${{\mathcal{H}}_2}$. 

\medskip

In a completely analogous way it can be justified that 
\begin{eqnarray}\label{ellder}
&& n^e\nabla_e \ell_b= \ell_b\left(n^f n^e\nabla_e \ell_f\right) + n_b\left(\ell^f n^e\nabla_e \ell_f\right)-m_b\left(\overline{m}^f n^e\nabla_e \ell_f\right) - 
\overline{m}_b\left(m^f n^e\nabla_e \ell_f\right) \nonumber \\ && \phantom{n^e\nabla_e \ell_b} = - \overline{\tau}\, m_b-\tau\,\overline{m}_b=-\overline{\tau}\, m'_b-\tau\,\overline{m}'_b+2\, u\, \tau\,\overline{\tau}\, n_b\,,
\end{eqnarray}
where in the last step the relation (\ref{mvesszo}), along with its complex conjugate, has been applied. Then, by referring to the $u$-independence of $\tau$ and to the fact that the vector fields $n^a, m'{}^a$ and $\overline{m}'{}^a$ are parallelly propagated it can be seen that 
\begin{equation}\label{ellvesszo}
\ell'{}^a=\ell^a+u\,\overline{\tau}\,m^a + u\,\tau\,\overline{m}{}^a + u^2\,\tau\,\overline{\tau}\, n^a\,
\end{equation}
is also parallelly propagated along the null generators of ${{\mathcal{H}}_2}$. It can also be verified that besides being parallelly propagated the system $\{\ell'{}^a,n^a,m'{}^a,\overline{m}'{}^a \}$ comprises a complex null tetrad along the null generators of ${{\mathcal{H}}_2}$. 

\bigskip

Notice also that the two complex null tetrads $\{\ell'{}^a,n^a,m'{}^a,\overline{m}'{}^a \}$ and $\{\ell^a,n^a,m^a,\overline{m}^a \}$ are related to each other by a tetrad rotation of class II, as given, e.g.\,on pages 53-55, in particular, by Eq.\,(346) of \cite{CH}, with $b=u\tau$. Thereby, the relation 
\begin{eqnarray}
&&\Psi _0'= \Psi _0 + 4 \,u\,\tau\, \Psi_1 + 6\, u^2\,\tau^2 \,\Psi_2 \label{psitr1}\\
&&\Psi _1' = \Psi _1 + 3\, u\,\tau\, \Psi_2\label{psitr2} \\ 
&&\phi _0' = \phi _0 + 2\, u\,\tau\, \phi_1 \label{psitr3}\,
\end{eqnarray}
holds for the transformed Weyl and Maxwell spinor components on ${{\mathcal{H}}_2}$ given with respect to $\{\ell'{}^a,n^a,m'{}^a,\overline{m}'{}^a \}$, where, in deriving (\ref{psitr1})-(\ref{psitr3}), the vanishing of $\Psi _3$, $\Psi _4$ and $\phi_2$ on ${{\mathcal{H}}_2}$ was also used.

Then, by making use of (\ref{Psi1n1}) and (\ref{Psi0n1}) $\Psi _0'=-C_{abcd}\ell'{}^am'{}^b\ell'{}^cm'{}^d $, $\Psi _1'=-C_{abcd}\ell'{}^an^b\ell'{}^cm'{}^d$ and $\phi _0'=F_{ab}\ell'{}^am'{}^b$ can be given on ${{\mathcal{H}}_2}$ as
\begin{eqnarray}
&&%&\hskip-.1cm
{\Psi}'_0=\frac{1}{2}\,u^2\left[\delta^2\Psi_2 + (\tau-2\,\beta)\,\delta\Psi_2 + 2\,k\left[\delta^2\phi_1-\left(\tau+2\,\beta\,\right)\delta\phi_1\right]\overline{\phi}_1+ k\,(\delta{\phi_1})\,\delta\overline{\phi}_1 \right]\\
&&%&\hskip-.1cm
{\Psi}'_1=u\,\left[\delta\Psi_2 +k\,(\delta\,\phi_1)\,\overline{\phi}_1\right]\\ 
&&%&\hskip-.1cm
{\phi}'_0=u\,\delta\phi_1 \,,
\end{eqnarray}
which completes the proof. \hfill\fbox{}\bigskip

\smallskip

Note that the potential occurrence of parallelly propagated curvature blow up in stationary black hole spacetimes, the event horizon of which is a Killing horizon, was already noted in \cite{r1} (see Remark 6.2 there for more details).  We would like to emphasize that the unbounded growth of the curvature tensor in parallelly propagated tetrads is novel in the sense that the curvature blow up, as opposed to spacetime singularities (see, e.g. \cite{HE,schmidt,wald} for more details) has nothing to do with geodesic incompleteness as the blow up occurs at the asymptotic ends of geodesically complete null generators of ${{\mathcal{H}}_1}$ and ${{\mathcal{H}}_2}$, i.e., infinitely far, in affine distance, from the points of ${{\mathcal{H}}_1}$ and ${{\mathcal{H}}_2}$. 

\medskip

It is also important to emphasize that---as it will be shown in Section \ref{sym} below---the found parallelly propagated curvature blow up is weak in the sense that neither of the scalar invariants, that can be built up from the Ricci and Weyl tensors, blows up. 

\section{On the existence of the horizon Killing vector field}\label{sym}
\renewcommand{\theequation}{6.\arabic{equation}}
\setcounter{equation}{0}

This section is to justify that non-trivial (global) one-parameter group of isometries can be associated with the selected spacetimes. 

\begin{theorem}\label{symmetry}
Suppose that $(M,g_{ab})$ is a four-dimensional electrovacuum spacetime with Maxwell field $F_{ab}$, allowing a non-zero cosmological constant, and possessing a pair of null hypersurfaces $\mathcal{H}_1$ and $\mathcal{H}_2$ generated by expansion and shear free geodesically complete null congruences such that they intersect on a two-dimensional spacelike surface, $\mathcal{Z}=\mathcal{H}_1\cap\mathcal{H}_2$. Then, there exist a non-trivial Killing vector field $K^a$ on the domain of dependence $D[{\mathcal{H}}_1\cup {\mathcal{H}}_2]$ such that 
\begin{itemize}
\item[(i)]  $K^a$ is unique up to a constant rescaling\,,
\item[(ii)]$K^a$ is tangent to the null generators of $\mathcal{H}_1$ and $\mathcal{H}_2$\,, 
\item[(iii)] we have that $\pounds_K F_{ab}=0$ on $D[{\mathcal{H}}_1\cup {\mathcal{H}}_2]$\,.
\end{itemize}
\end{theorem} 
\noindent\textbf{Proof:}{\ } The justification of the above assertions can be given based on the proof of Proposition B.1. of \cite{frw}. To see this, note first that the gauge fixings applied in the present paper are identical to those specified on page 702 in \cite{frw}, and only some minor modifications are needed (as indicated below) in consequence of the inclusion of a non-zero cosmological constant $\widetilde\Lambda$.   

\medskip

By referring Lemma B.3. of \cite{frw} a to be non-trivial Killing vector field $K^a$ on $D[{\mathcal{H}}_1\cup {\mathcal{H}}_2]$ can be seen to possess, up to a constant factor, the form
\begin{equation} 
{K}^a= \left\{ 
\begin{array} {r l}  
\hskip-0.5cm{-}\,r\,\ell^a\,\,, & {\rm on}\ \ \mathcal{H}_1\,,\\ 
\hskip-0.5cm\phantom{-,}\,\,u\, n^a\,\,, & {\rm on}\ \ \mathcal{H}_2\,. 
\end{array}
\right.\label{KVF}
\end{equation}
on our initial data surface comprised by $\mathcal{H}_1$ and $\mathcal{H}_2$.

\medskip

It is straightforward to verify that the inclusion of a non-zero cosmological constant $\widetilde\Lambda=6 \Lambda$ implies, first of all, that the functional form of $\mu$, given by (B.2) in \cite{frw}, on $\mathcal{H}_1$ and $\rho$, given by (B.3) in \cite{frw}, on $\mathcal{H}_2$ have to be replaced by (\ref{mu}) and (\ref{rhon2}), respectively. Note that the justification of the vanishing of the Lie derivatives of the metric on the null hypersurfaces $\mathcal{H}_1$ and $\mathcal{H}_2$ in the second part of the proof of Lemma B.6 of \cite{frw} is not affected by the inclusion of the constant $\Lambda$. In addition, there is also a minor modification of (B.11) \cite{frw} where the term $-\widetilde\Lambda\,\pounds_K g_{ij}$ appears on the right hand side in the present case. Note however, that this does not affect the argument in Lemma B.6 as the pertinent system (B.10), (B.12) and (B.13) will comprise  a homogeneous linear system for the unknowns $\pounds_K g_{ij}$ and $\pounds_K F_{ij}$. The rest of 
the proof goes through without further 
modifications.   
\hfill\fbox{}\bigskip

\bigskip

Note that by property $(ii)$ the null hypersurfaces $\mathcal{H}_1$ and $\mathcal{H}_2$ are invariant under the action of the one-parameter group of isometries associated with the Killing vector field $K^a$. In addition, the explicit form (\ref{KVF}) of the Killing vector field $K^a$ verifies that  $K^a$ is null on $\mathcal{H}_1$ and $\mathcal{H}_2$ which justifies that the null hypersurfaces $\mathcal{H}_1$ and $\mathcal{H}_2$ are Killing horizons, respectively \cite{rw1,rw2}. As a bifurcate Killing horizon $\mathcal{H}$ is comprised by two Killing horizons which intersect on a spacelike two-dimensional surface, called the bifurcation surface of $\mathcal{H}$, we have
\begin{corollary}
The pair of null hypersurfaces $\mathcal{H}_1$ and $\mathcal{H}_2$ comprise a bifurcate type Killing horizon with bifurcation surface $\mathcal{Z}$. 
\end{corollary}

Recall that in modeling generic stationary distorted black hole spacetimes it is usually assumed that they possess only a single (global) one-parameter group of isometries associated with a Killing horizon. Note that beside assuming the existence of the null hypersurfaces $\mathcal{H}_1$ and $\mathcal{H}_2$ comprising the bifurcate type Killing horizon, generated by complete null geodesics, no assumption concerning the asymptotic structure has been made. Since, the freely specifiable data on the bifurcation surface  $\mathcal{Z}$ need not to have any symmetry, by the results covered in \cite{rkill,rkill2}, the pertinent spacetimes do not either possess any symmetry in addition to the horizon Killing vector field. Thereby, there is a one-to-one correspondence between the class of spacetimes selected in this paper and the class of generic stationary distorted electrovacuum black hole spacetimes. 

\medskip

Let us now return to our assertion concerning the strength of the parallelly propagated curvature blow up. Such a blow up is considered to be strong whenever either of the scalar invariants that can be built up from the Ricci and Weyl tensors, also blows up along the pertinent null generator of ${{\mathcal{H}}_1}$ or ${{\mathcal{H}}_2}$. That this does not happen can be verified by making use of the existence of the horizon compatible Killing vector field $K^a$. Since the Lie derivative of the spacetime metric, and in turn that of the Ricci and Weyl tensor, must be identically zero with respect to $K^a$, any sort of scalar curvature expression constructed from these tensors and the metric has to be also invariant along the generators of ${{\mathcal{H}}_1}$ and ${{\mathcal{H}}_2}$ to which $K^a$ is tangential. This justifies 
\begin{lemma}
Neither of the parallelly propagated curvature blow up is strong, i.e., the scalar invariants that can be built up from the Ricci and Weyl tensors remain constant along the generators of ${{\mathcal{H}}_1}$ and ${{\mathcal{H}}_2}$. 
\end{lemma}

In virtue of this result there is an immediate need to clear up what is then the significance of the parallelly propagated curvature blow up along the generators of ${{\mathcal{H}}_1}$ and ${{\mathcal{H}}_2}$. In this respect it is informative to recall that the Killing vector field $K^a$, as given by (\ref{KVF}), itself has a parallelly propagated blow up along the generators of ${{\mathcal{H}}_1}$ and ${{\mathcal{H}}_2}$ with respect to a complex null tetrads $\{\ell^a,n^a,m^a,\overline{m}^a \}$ and $\{\ell'{}^a,n^a,m'{}^a,\overline{m}'{}^a \}$, respectively. Thereby the invariance of the Ricci and Weyl tensors with respect to coordinate basis fields, adopted to the Killing vector field $K^a$, is not at all in conflict with the blow up of these tensors with respect to parallelly propagated tetrads (see also the Proposition in Section 3 of \cite{rw1}). {Based on these observations the parallelly propagated curvature blow up could simply be interpreted as a redshift effect. Nevertheless, we would 
like to} emphasize that the occurrence of a parallelly propagated curvature blow up implies that some of the physically measurable quantities, e.g.\,the tidal force tensor components of the curvature tensor, associated with the null congruences transverse to ${{\mathcal{H}}_1}$ and ${{\mathcal{H}}_2}$, get{s to be larger and larger} while approaching the future and past ends of some of the null generators of ${{\mathcal{H}}_1}$ and ${{\mathcal{H}}_2}$.

\section{The near Schwarzschild black holes}\label{sing2}
\setcounter{equation}{0}

This section is to explore the near Schwarzschild solutions in the space of stationary distorted vacuum black hole spacetimes and also to justify our claim about the universal occurrence of parallelly propagated curvature blow up.

\medskip

Consider first the topology of the cross section $\mathcal{Z}$. For the shake of simplicity we shall assume that it has the topology of a two-sphere. Note that---in virtue of  Hawking's black hole topology theorem \cite{hawk} (see also \cite{ga1,rnew})---this assumption holds for all the considered distorted electrovacuum black hole configurations.

\medskip

Consider now the two-metric $g_{AB}$ on $\mathcal{Z}$. It is said that a metric $g_{AB}$ is conformal equivalent to the unit sphere metric $g_{AB}^\circ$ if there exist a diffeomorphism $\varphi : \mathcal{Z} \rightarrow {\mathcal{Z}}$ and a positive smooth function $\Omega$ such that
\begin{equation}
g_{AB}=\Omega^2 \hskip-.1cm\cdot\hskip-.05cm\varphi^*g_{AB}^\circ\,.
\end{equation}
If the diffeomorphism $\varphi$ is the identity then $g_{AB}$ and $g_{AB}^\circ$ are called to be conformally related. It is well-known that such a diffeomorphism---that can be given in terms of suitable choice of new coordinates---always exists. Thereby, we shall assume that the metric $g_{AB}$ on $\mathcal{Z}$ is conformally related to the unit sphere metric $g_{AB}^\circ$, i.e.\,it will be assumed that the diffeomorphism $\varphi: \mathcal{Z} \rightarrow{\mathcal{Z}}$ has already be performed.
%\,\footnote{The conformal class of the unit two-sphere metric $g_{AB}^\circ$ consists of those metrics on $\mathcal{Z}$ that are conformally related to $g_{AB}^\circ$. Thereby, $g_{AB}$ belongs to this conformal class.} 

\medskip

Assume that $(\theta,\phi)$ are corresponding distinguished spherical coordinates on $\mathcal{Z}$, and define the holomorphic coordinate $\xi$ there as 
\begin{equation}\label{xiA}
\xi=\log\left[\tan\frac{\theta}{2}\right]+i\,\phi\,.
\end{equation}
Then the complex vector field $\xi^A$ may be given as 
\begin{equation}\label{xiAv}
\xi^A=-\frac{\sqrt{2}}{2}\,{\overline P}\left(\partial_\xi\right)^A=-\frac{\sqrt{2}}{2}\,{\overline P}\left[{\sin\theta}\,\left(\partial_\theta\right)^A+i\,\left(\partial_\phi\right)^A \right]\,,
\end{equation}
where $P$ is a non-vanishing but otherwise arbitrary smooth complex valued function on $\mathcal{Z}$. By making use of the dual, 
\begin{equation}
\xi_A=-\frac{\sqrt{2}}{2\overline P} \left(d\xi\right)_A=-\frac{\sqrt{2}}{2\overline P}\left[\frac{(d\theta)_A}{\sin\theta}+i\,(d\phi)_A \right]\,,
\end{equation}
of $\xi^A$, the covariant form of the metric $g_{AB}$ on $\mathcal{Z}$ can be given as 
\begin{equation}
g_{AB}=-\frac{1}{P\overline P}\left(d\xi\right)_{(A}\left(d\overline \xi\right)_{B)}=
-\frac{1}{P\overline P \sin^2\theta}\left[(d\theta)_A (d\theta)_B +\sin^2\theta\, (d\phi)_A (d\phi)_B\right]\,.
\end{equation}
Note that, according to our assumptions, this metric is conformally related to the unit sphere metric $g_{AB}^\circ=(d\theta)_A (d\theta)_B +\sin^2\theta\, (d\phi)_A (d\phi)_B$ with the positive smooth function 
\begin{equation}
\Omega=\frac{1}{\sqrt{P\overline P} \sin\theta}\,.
\end{equation}

\medskip

As $\mathcal{Z}$ has the topology of a two-sphere the `complex curvature' ${K}$ and the Gaussian curvature $\mathcal{K}_G$ of $g_{AB}$ can be given (see (4.14.20) and Proposition 4.14.21. of  \cite{penrose:rindler}) as 
\begin{equation}\label{komlexcurv}
{K}=-\Psi_2 +\rho\,\mu - \sigma\,\lambda + k\,\phi_1\overline\phi_1 +\Lambda\,%.
\end{equation}
and 
\begin{equation}\label{psi2gauss}
\mathcal{K}_G={K}+\overline{K}\,,
\end{equation}
respectively. 

\medskip

Then the following characterization of the Schwarzschild black holes can be given. 

\begin{proposition}\label{schw}
For the Schwarzschild solution no parallelly propagated curvature blow up occurs.
\end{proposition}
\noindent\textbf{Proof:}{\ } Note first that as the Schwarzschild solution is spherically symmetric it is natural to choose $\xi^A$ on $\mathcal{Z}$ as
\begin{equation}\label{xiA_sch}
\xi^A=\frac{\sqrt{2}}{2 r_\circ}\left[\left(\partial_\theta\right)^A+\frac{i}{\sin\theta}\,\left(\partial_\phi\right)^A \right]\,,
\end{equation}
where $r_\circ$ denotes the radius of the ball representing the bifurcation surface $\mathcal{Z}$. Then, the metric on $\mathcal{Z}$ reads as 
\begin{equation}\label{metric_sch}
g_{AB}=-r_\circ^2\left[(d\theta)_A (d\theta)_B +\sin^2\theta\, (d\phi)_A (d\phi)_B\right]\,.
\end{equation}
As the Schwarzschild spacetime is a vacuum solution the rest of the freedom we have in specifying a reduced initial data set $\mathbb{V}^{red}_0$ is exhausted by choosing $\tau=0$ on $\mathcal{Z}$. 

Then, by solving (\ref{beta}) we get 
\begin{equation}
\beta=\frac{\sqrt{2}\cos\theta}{4\,r_\circ\sin\theta}\,,
\end{equation}
which, along with the relation $\alpha=\overline{\tau}-\overline{\beta}$ and (\ref{psi2}), yields that 
\begin{equation}\label{psi2sch}
\Psi_2=-\frac{1}{2\,r_\circ^2}\,,
\end{equation}
on $\mathcal{Z}$. 

\medskip

Clearly, then all the expressions $\tau$, $\delta\Psi_2$, $\delta^2\Psi_2$, $\overline{\delta}\Psi_2$ and $\overline{\delta}^2\Psi_2$ vanish throughout $\mathcal{Z}$, which along with (\ref{Psi3}), (\ref{Psi4}),  (\ref{ndl1}), (\ref{ndl2}) and Proposition\,\ref{ndl}, implies that no parallelly propagated blow up of the Weyl tensor occurs along the generators of ${\mathcal{H}}_1$ and ${\mathcal{H}}_2$ in the corresponding spacetimes. 

What remained to be justified is that the above choice of $\xi^A$ and $\tau$ is compatible with the Schwarzschild solution. In doing so we shall need the following
\begin{lemma}\label{schwl}
For the Schwarzschild solution the spin coefficients $\tau,\sigma,\nu,\lambda$ and the Weyl spinor components $\Psi_0, \Psi_1, \Psi_3, \Psi_4$ are identically zero on the domain of dependence $D[{\mathcal{H}}_1\cup {\mathcal{H}}_2]$.
\end{lemma}
\noindent\textbf{Proof:}{\ } Let us start be verifying first that the spin coefficients $\tau,\sigma,\nu,\lambda$ and the Weyl spinor components $\Psi_0, \Psi_1, \Psi_3, \Psi_4$ identically vanish on the initial data surface ${\mathcal{H}}_1\cup {\mathcal{H}}_2$.  

To see that this is the case notice first the by (\ref{sigma}) and by the vanishing of $\tau$ we have that $\sigma=0$ on ${{\mathcal{H}}_2}$. Similarly, by (\ref{Psi1n1}) and (\ref{Psi0n1}), along with the vanishing of $\tau$, $\delta\Psi_2$ and $\delta^2\Psi_2$, the latter relations follow from (\ref{psi2sch}), we also have that both $\Psi_0$ and $\Psi_1$ vanish on ${{\mathcal{H}}_2}$. Since $\tau,\sigma$ and $\Psi_0, \Psi_1$ identically zero on ${{\mathcal{H}}_1}$ we already have justified half of the relations involved our claim concerning the initial data on ${\mathcal{H}}_1\cup {\mathcal{H}}_2$. 

The vanishing of the spin coefficients $\nu,\lambda$ and the Weyl spinor components $\Psi_3, \Psi_4$ on the initial data surface ${\mathcal{H}}_1\cup {\mathcal{H}}_2$ can be justified analogously by making use of (\ref{nun1}), (\ref{psi3n2}) and (\ref{psi4n2}). 

Finally, the assertion of our lemma follows then from the fact that the vacuum version of the equations (EM.6-7), (EM.11), (EM.13-15) and (EM.17-18) comprise a first order symmetric hyperbolic system that is homogeneous and linear in the variables $\tau,\sigma,\nu,\lambda$ and $\Psi_0, \Psi_1, \Psi_3, \Psi_4$. As these type of partial differential equations possess unique solutions they have the identically zero solution for vanishing initial data.  
\hfill\fbox{}\bigskip

Returning to our proof of the proposition, note that by the vanishing of the spin coefficients $\tau,\sigma,\nu,\lambda$ and the Weyl spinor components $\Psi_0, \Psi_1, \Psi_3, \Psi_4$, and by making use of the reduced equations (EM.1-5), (EM.8-10), (EM.12) and $(NP.6.12b)$---these are radial ordinary differential equations along the null geodesics transverse to ${{\mathcal{H}}_2}$ with tangent $\ell^a$---, all the variables in $\mathbb{V}$ can be determined not only on $D[{\mathcal{H}}_1\cup {\mathcal{H}}_2]$ but everywhere in $\mathcal{O}$, where the Gaussian null coordinates are defined. In particular, we have that  the relations
\begin{eqnarray}
\xi^A(u,r)&&\hskip-.6cm=\frac{\xi^A}{1+ u\,r\,\Psi_2} \label{xiAsch} \\
\omega(u,r)&&\hskip-.6cm=0\\
X^A(u,r)&&\hskip-.6cm=0\\
U(u,r)&&\hskip-.6cm=-\frac{r^2\,\Psi_2}{1+ u\,r\,\Psi_2} \label{Usch} %\\
%\rho(u,r)&&\hskip-.6cm=-\frac{u\,\Psi_2}{1+ u\,r\,\Psi_2} \\
%\gamma(u,r)&&\hskip-.6cm=\frac{1}{2\,u} \left[1-\left(1+ u\,r\,\Psi_2\right)^{-2}\right] 
\end{eqnarray}
hold in $\mathcal{O}$. In (\ref{xiAsch})-(\ref{Usch}) $\xi^A$ and $\Psi_2$ with suppressed arguments denote the restrictions of these variables onto $\mathcal{Z}$, and thus they are independent of $u$ and $r$. 

The justification that these functions do really uniquely determine the Schwarzschild solution may be done in two steps. Restrict first attention to $\mathcal{O}^+\subset\mathcal{O}$ comprised by points subject to the relations $0 < u < \infty$ and $-\infty < r\,u < 2\,r_\circ^2$.\,\footnote{Note that a similar construction works for $\mathcal{O}^-\subset\mathcal{O}$ consisting of points with $-\infty<u<0$ and $-2\,r_\circ^2 < r\,u < \infty$.} Perform then a rescaling of the null vector fields $\ell^a\rightarrow \widetilde\ell^a=A^{-1}\ell^a$, $n^a\rightarrow \widetilde n^a=A\,n^a$, with $A=\frac{u}{2\,r_\circ}$. This induces in $\mathcal{O}^+$ (see, e.g., \cite{rw1}) a replacement of the Gaussian null coordinates $u$ and $r$ by $\widetilde u$ and $\widetilde r$ with 
\begin{equation}\label{ur}
u=\exp\left(\frac{\widetilde u}{2\,r_\circ}\right)\,,\hskip0.5cm r=\frac{{2\,r_\circ}\widetilde r}{u}\,.
\end{equation}

Second, introduce the Schwarzschild type radial coordinate, $0<\widehat r<\infty$ in $\mathcal{O}^+$, determined by the relation $\widehat r=r_\circ-\widetilde r$. Then, in virtue of (\ref{psi2sch}) and (\ref{ur}), it is straightforward to check that 
\begin{equation}
{1+ u\,r\,\Psi_2}=\frac{\widehat r}{r_\circ}\,,
\end{equation}
which along with (\ref{xiAsch}) implies that throughout $\mathcal{O}^+$ the metric $g_{AB}=-(\xi_A\overline\xi_B+\overline\xi_A\xi_B)$ is conformal to the unite sphere metric with conformal factor $\widehat r^2$. Similarly, it can be verified that 
\begin{equation}%\label{psi2sch}
g^{\widehat r\widehat r}=g^{ab}(d\widehat r)_a(d\widehat r)_b=g^{ab}(d\widetilde r)_a(d\widetilde r)_b=\left(\frac{ u}{2\,r_\circ}\right)^2g^{rr}+\left(\frac{u\,r}{2\,r_\circ^2}\right)\,g^{ur}=\frac{r_\circ}{\widehat r}-1\,,
\end{equation}
as it should hold for the Schwarzschild solution in $\mathcal{O}^+$ in ingoing null coordinates $(\widetilde u,\widehat r,\theta,\phi)$ and with $r_\circ=2 M$.
\hfill\fbox{}\bigskip

We have seen that both $\delta\Psi_2$ and $\overline{\delta}\Psi_2$, along with $\tau$, vanish on $\mathcal{Z}$ for the Schwarzschild solution.  As it is justified by the following result that these conditions are not only necessary but also sufficient.

\begin{proposition}\label{charSch}
The simultaneous and identical vanishing of $\delta\Psi_2$, $\overline{\delta}\Psi_2$, $\delta\phi_1$ and $\overline{\delta}\phi_1$ on $\mathcal{Z}$ implies that $\mathcal{Z}$ is a metric sphere. In particular, the simultaneous and identical vanishing of $\delta\Psi_2$, $\overline{\delta}\Psi_2$ and $\tau$ on $\mathcal{Z}$ uniquely single out the Schwarzschild solution within the set of stationary distorted vacuum black hole spacetimes.
\end{proposition}
\noindent\textbf{Proof:}{\ } 
Note that, in virtue of (\ref{dop}) and (\ref{xiAv}), the simultaneous vanishing of $\delta\Psi_2$ and $\overline{\delta}\Psi_2$, and similarly, the simultaneous vanishing of $\delta\phi_1$ and $\overline{\delta}\phi_1$, imply that both $\Psi_2$ and $\phi_1$ have to be constant throughout $\mathcal{Z}$. 
Taking then into account (\ref{psi2}), (\ref{komlexcurv}) and (\ref{psi2gauss}), along with the vanishing of $\rho, \sigma, \mu$ and $\lambda$ on $\mathcal{Z}$ we get that the Gaussian curvature 
\begin{equation} 
\mathcal{K}_G=-\,\left[\Psi_2+\overline{\Psi}_2-2\,\Lambda - 2\,k\,\phi_1\,\overline{\phi}_1\right]\,
\label{psi2_2}  
\end{equation}
has to be constant throughout $\mathcal{Z}$, which justifies that $\mathcal{Z}$ has to be a metric sphere, i.e., the metric $g_{AB}$ on  $\mathcal{Z}$  is of the form (\ref{metric_sch}). 

\medskip

In justifying the second part of our statement, note that by making use of an appropriate coordinate transformation of the form $x^A\rightarrow \widetilde x^A=\widetilde x^A(x^3,x^4)$ on ${\mathcal{Z}}$ (as described in Section\,\ref{stac}) the vector field $\xi^A$ on ${\mathcal{Z}}$ can be put into the form (\ref{xiA_sch}) with $r_\circ=\frac{1}{\sqrt{\mathcal{K}_G}}$. Thereby, in virtue of Proposition\,\ref{schw} we have that the simultaneous vanishing of $\delta\Psi_2$,  $\overline{\delta}\Psi_2$ and $\tau$ uniquely determine the Schwarzschild solution. 
\hfill\fbox{}\bigskip

Consider now a stationary distorted black hole with $\tau=0$ on $\mathcal{Z}$ but for which either $\delta\Psi_2$ or $\overline{\delta}\Psi_2$ is non-zero somewhere on $\mathcal{Z}$. Note that whenever $\tau=0$ on $\mathcal{Z}$, in virtue of (\ref{psi2}), (\ref{komlexcurv}) and (\ref{psi2gauss}), in the vacuum case with $\Lambda=0$, $\Psi_2$ is real and $\mathcal{K}_G=-2\,\Psi_2$. Therefore such a configuration cannot belong to the Schwarzschild family. Since then either ${\widetilde\Psi_1}'=\delta\Psi_2$ or $\widetilde\Psi_3=\overline{\delta}\Psi_2$ must be non-zero somewhere on $\mathcal{Z}$, in virtue of Proposition\,\ref{ndl}, we infer that a parallelly propagated blow up of the Weyl tensor has to occur. More definitely, the contractions $\Psi _1'=-C_{abcd}\ell'{}^an^b\ell'{}^cm'{}^d$ and  $\Psi _3=-C_{abcd}n^a\ell^b n^c\overline{m}^d $ tend to infinity---, while approaching the future and past ends of some of the null generators of ${{\mathcal{H}}_1}$ and 
$\mathcal{H}_2$. Since this curvature blow up occurs regardless how small is the deviation of $\Psi_2$ from a constant value we have the following 

\begin{corollary}
Consider the space of vacuum solutions to Einstein's equations. Then, in an arbitrarily small neighborhood of the Schwarzschild solution there always exist distorted vacuum black hole configurations such that parallelly propagated curvature blow up occur along some of the null generators of their bifurcate Killing horizon both to the future and to the past.
\end{corollary}

After having all the above characterization of the Schwarzschild and `nearly' Schwarzschild solutions in the space of stationary distorted black hole spacetimes it is natural to ask what can be said about the Kerr{-Newman} black hole solutions. Recall that the Gaussian curvature $\mathcal{K}_G$ of the bifurcation surface of {a non-extremal} Kerr-Newman solution{ with non-zero specific angular momentum parameter $a$}\,\footnote{{In case of positive cosmological constant the Gauss curvature of the bifurcation surface of a non-extremal event or cosmological horizon of a Kerr-Newman-de Sitter spacetime may also be constant (see e.g.\,equation (3.6) of \cite{engman}) if $\Lambda=3/r_H^2$ where $r_H$ denotes the radius of either the event or the cosmological horizon.}} is known to be non-constant \cite{smarr}. Thereby, in virtue of  (\ref{komlexcurv}) and (\ref{psi2gauss}), the simultaneous vanishing of both $\delta(\Psi_2+\overline{\Psi}_2{-2\,k\,\phi_1\,\overline{\phi}_1-2\,\Lambda})$ and $\overline{\delta}(\
Psi_2+\overline{\Psi}_2{-2\,k\,\phi_1\,\overline{\phi}_1-2\,\Lambda})$ can occur only at the poles and at the equatorial of $\mathcal{Z}$. This implies then that either $\delta\Psi_2$, $\overline{\delta}\Psi_2${, $\delta\phi_1$ or $\overline{\delta}\phi_1$} must be non-zero everywhere else. This, in 
virtue of the above observations, justifies 

\begin{corollary}
Parallelly propagated curvature blow up occur both to the future and to the past ends along almost all of the null generators of the bifurcate Killing horizon of {non-extremal} Kerr-Newman spacetimes{ with non-zero specific angular momentum parameter}.
\end{corollary}

Regardless of the occurrence of these parallelly propagated curvature blow up it would be important to establish the correspondent of Proposition\,\ref{charSch}, i.e.\,to know what are the minimal geometrical conditions that distinguishes the Kerr-Newman family within the set of generic stationary distorted electrovacuum black hole spacetimes. 

\section{Final remarks}\label{con}
\setcounter{equation}{0}

In this paper {first a systematic investigation of the generic null characteristic initial value problem within the setup of Newman-Penrose formalism was given for smooth four-dimensional electrovacuum spacetimes allowing non-zero cosmological constant. Then, based on the yielded results, a detailed investigation of generic stationary distorted electrovacuum black hole configurations was also given}. We would like to emphasize
again that while all the previous investigations related to distorted
black hole spacetimes were restricted (almost) exclusively to the
static axially symmetric vacuum solutions the geometrical
framework introduced in this paper is  suitable to investigate all the
possible stationary distorted electrovacuum black hole configurations. 
{Note also that even this application could not be done without properly separating a suitable reduced subsystem of evolution equations, in the generic setup, from the coupled Newman-Penrose and Maxwell equations.} 

\medskip

Our main result justifies that the geometry of any four-dimensional electrovacuum distorted black hole is uniquely determined. Once a complex vector field $\xi^A$ (determining the metric induced on the bifurcation surface), the $\tau$ spin coefficient and the $\phi_1$ electromagnetic potential are specified at the bifurcation surface. In this respect it seems to be quite appropriate to think of the bifurcation surface $\mathcal{Z}=\mathcal{H}_1\cap\mathcal{H}_2$ of such a generic four-dimensional stationary distorted electrovacuum black hole spacetime as the unique compact ``carrier'' of the initial data. In other words, we may think of the bifurcation surface as a ``holograph'' storing all the basic information which, along with the field equations, can be used to reconstruct the entire four-dimensional stationary distorted electrovacuum black hole spacetime.   

Note that the well-known asymptotically flat or asymptotically (locally) anti-de-Sitter stationary electrovacuum black hole spacetimes---distinguished by the black hole uniqueness theorems---do belong to the set of distorted  black hole spacetimes. However, this set is significantly larger than that of the  asymptotically flat or asymptotically (locally) anti-de-Sitter stationary electrovacuum black hole spacetimes as, besides the existence of the pair of null hypersurfaces $\mathcal{H}_1$ and $\mathcal{H}_2$, no assumption on the asymptotic structure was made. 

It is well-known that all the asymptotically flat or asymptotically (locally) anti-de-Sitter stationary electrovacuum black hole spacetimes are both analytic and special in their algebraic type. Therefore one would expect that the following construction---analogous to the one applied above in deducing (\ref{xiAsch})-(\ref{Usch}) in the Schwarzschild case---could be applied. One should start by the data specified at the bifurcation surface and then integrate the field equations along suitably chosen subfamilies of the principal null congruences. 
In this respect it would be important to find the precise conditions singling out the rotating asymptotically flat or asymptotically (locally) anti-de-Sitter black hole configurations in the set of freely specifiable functions at the bifurcation surface.

Obviously, the identification of these selection rules would open a completely new avenue in the black hole uniqueness problem already in case of four-dimensional spacetimes. Note, however, that the success of the associated program should be even more far-reaching. Concerning the huge variety of stationary black hole configurations in higher dimensional theories it would be of great importance to work out generalizations of the results obtained here. As some of the fundamental techniques applied in this paper have already been generalized to higher dimensions the identification of the appropriate selection rules would offer significant insight into machinery of the higher dimensional black hole uniqueness problem. The investigation of these and related issues would definitely deserve further attention.

\medskip

Perhaps the most interesting and somewhat unexpected result of the present paper is related to the universal occurrence of parallelly propagated curvature blow up along some of the generators of the bifurcate Killing horizon of stationary distorted electrovacuum black hole spacetimes. Clearly, it would be of great importance to know whether these results have any implications in connection with the stability of the well-known asymptotically flat or asymptotically (locally) anti-de-Sitter stationary electrovacuum black hole spacetimes. 

\medskip

There are some immediate similarities between this latter result and the outcome of the recent investigations of the evolution of massless scalar Klein-Gordon field on extremal Reissner-Nordstr\"om or Kerr backgrounds \cite{aret1,aret2,aret3} (for some additional explanatory investigations see also \cite{sergio,fridbiz,lucietti}). Therefore, it is of obvious interest to relate them. In \cite{aret3} Aretakis showed that instabilities of solutions to the wave equation develop asymptotically along the event horizon of fixed stationary axisymmetric extremal black hole backgrounds. The blow up of second or third order transverse derivatives of the scalar field along the null generators of the horizon is certainly analogous to the parallelly propagated curvature blow up we have found. Aretakis also argued that only local properties of the black hole spacetime came into play. This is also in accordance to our findings as we have full control of geometric and physical quantities only on the horizon. In addition, in 
both 
cases the blow up occur to the asymptotic end(s) of the horizon generators. There are, however, significant discrepancies, as well. While in our case the occurrence of the curvature blow up is shown to be a universal property of {\it non-extremal} distorted black hole solutions to the fully non-linear Einstein-Maxwell equations, Aretakis' results hold for extremal black hole configurations exclusively and they merely guarantee the blow up of certain higher order transverse derivatives of a linear test field. It would be interesting to know whether the latter also indicates the occurrence of some sort of parallelly propagated blow up of certain derivatives of the energy-momentum tensor in the sense it is defined, e.g., in \cite{schmidt} (see also the proof of Proposition\,\ref{ndl} above). 

\medskip

\section*{Acknowledgments}

The author wishes to thank Robert Wald, Akihiro Ishibashi, Marek Rogatko and Mihalis Dafermos for useful
comments.
This research was supported in part
%s by OTKA grant K67942 and 
by the Die Aktion Österreich-Ungarn, Wissenschafts- und Erziehungskooperation grant 87öu16. The author would also like to thank the organizers of the workshop ``Dynamics of General Relativity: Black Holes and Asymptotics'' held in Ervin Schr\"odinger Institute, 10-21 December 2012, for the possibility to present some of the preliminary results covered by this paper.

\appendix
\section{Appendix}\label{Appendix A}
\renewcommand{\theequation}{A.\arabic{equation}}
\setcounter{equation}{0}

This appendix is to provide the proofs of Theorems \ref{unique} and \ref{equivalent}. 

\begin{theorem}\label{unique2} 
Denote by $\mathbb{V}_0$ a full initial data set, satisfying the ``inner'' Newman-Penrose and Maxwell equations on the initial data surface comprised by the pair of intersecting null hypersurfaces ${\mathcal{H}}_1$ and ${\mathcal{H}}_2$. Then, there exist a unique solution, $\mathbb{V}$, on the domain of dependence $D[{\mathcal{H}}_1\cup {\mathcal{H}}_2]$, to the reduced Einstein-Maxwell equations, (EM.1) - (EM.21), such that $\mathbb{V}\vert_{{{\mathcal{H}}_1\cup {\mathcal{H}}_2}}=\mathbb{V}_0$.
\end{theorem} 
{\sl Proof:} 
In justifying the above statement we show first that the reduced Einstein-Maxwell equations (EM.1)-(EM.21), do really form a determined first order system of partial differential equations (PDE) for our unknowns. 

In doing so notice first that equations (EM.1)-(EM.21), when written out in Gaussian null coordinates $(u,r,x^3,x^4)$ in ${\mathcal{O}}$, possess the structure 
\begin{equation}\label{eqV}
\mathbb{A}^\mu \cdot \partial_\mu \mathbb{V} + \mathbb{B}=0,
\end{equation}
where the matrices $\mathbb{A}^\mu$ and $\mathbb{B}$ smoothly depend on $\mathbb{V}$, along with its complex conjugate $\overline{\mathbb{V}}$. Moreover, it can also be seen that the matrices $\mathbb{A}^\mu$ are Hermitian, i.e., $\overline{\mathbb{A}}{}^\mu{}^T={\mathbb{A}}^\mu$ and the combination
$\mathbb{A}^\mu(\ell_\mu+n_\mu)$ is positive definite in ${\mathcal{O}}$.

The validity of the latter assertions follow from the fact that the coefficient matrices of  the derivative  operators $\rm{D},\Delta,\delta,\overline\delta$ in (EM.1)-(EM.21) have the form of $23\times23$ matrices given as
\begin{equation}\label{M1}
\hskip-0.2cm\mathbb{A}^{\rm{D}}={\small\left( {\begin{array}{c|c}
\mathbf{1} & \mathbf{0}\\ \hline 
\mathbf{0} &  \begin{array}{rrrrrrrr}
 0 & 0 & 0 & 0 & 0 & 0 & 0 & 0 \\
 0 & 1 & 0 & 0 & 0 & 0 & 0 & 0 \\
 0 & 0 & 1 & 0 & 0 & 0 & 0 & 0 \\
 0 & 0 & 0 & 1 & 0 & 0 & 0 & 0 \\
 0 & 0 & 0 & 0 & 1 & 0 & 0 & 0 \\
 0 & 0 & 0 & 0 & 0 & 0 & 0 & 0 \\
 0 & 0 & 0 & 0 & 0 & 0 & 1 & 0 \\
 0 & 0 & 0 & 0 & 0 & 0 & 0 & 1
\end{array} 
\end{array}}
 \right) }
%\end{equation}
%\begin{equation}
\ \ \ \ \ 
\mathbb{A}^{\Delta}={\small\left( {\begin{array}{c|c}
\mathbf{0} & \mathbf{0}\\ \hline 
\mathbf{0} &  \begin{array}{rrrrrrrr}
 1 & 0 & 0 & 0 & 0 & 0 & 0 & 0 \\
 0 & 1 & 0 & 0 & 0 & 0 & 0 & 0 \\
 0 & 0 & 1 & 0 & 0 & 0 & 0 & 0 \\
 0 & 0 & 0 & 1 & 0 & 0 & 0 & 0 \\
 0 & 0 & 0 & 0 & 0 & 0 & 0 & 0 \\
 0 & 0 & 0 & 0 & 0 & 1 & 0 & 0 \\
 0 & 0 & 0 & 0 & 0 & 0 & 1 & 0 \\
 0 & 0 & 0 & 0 & 0 & 0 & 0 & 0 
\end{array} 
\end{array}}
 \right) }
\end{equation}
\begin{equation}\label{M2}
\hskip-0.1cm\mathbb{A}^{\delta}={\small\left( {\begin{array}{c|c}
\mathbf{0} & \mathbf{0}\\ \hline 
\mathbf{0} &  \begin{array}{rrrrrrrr}
 0 & \hskip-0.21cm-1 & 0 & 0 & 0 & 0 & 0 & 0 \\
 0 & 0 & \hskip-0.21cm-1 & 0 & 0 & 0 & 0 & 0 \\
 0 & 0 & 0 & \hskip-0.21cm-1 & 0 & 0 & 0 & 0 \\
 0 & 0 & 0 & 0 & \hskip-0.21cm-1 & 0 & 0 & 0 \\
 0 & 0 & 0 & 0 & 0 & 0 & 0 & 0 \\
 0 & 0 & 0 & 0 & 0 & 0 & \hskip-0.21cm-1 & 0 \\
 0 & 0 & 0 & 0 & 0 & 0 & 0 & \hskip-0.21cm-1 \\
 0 & 0 & 0 & 0 & 0 & 0 & 0 & 0 \\
\end{array} 
\end{array}}
 \right) }
%\end{equation}
%\begin{equation}\label{M3}
\ \ \ \ \ 
\hskip-0.05cm\mathbb{A}^{\overline\delta}={\small\left( {\begin{array}{c|c}
\mathbf{0} & \mathbf{0}\\ \hline 
\mathbf{0} &  \begin{array}{rrrrrrrr}
 0 & 0 & 0 & 0 & 0 & 0 & 0 & 0 \\
 \hskip-0.21cm-1 & 0 & 0 & 0 & 0 & 0 & 0 & 0 \\
 0 & \hskip-0.21cm-1 & 0 & 0 & 0 & 0 & 0 & 0 \\
 0 & 0 & \hskip-0.21cm-1 & 0 & 0 & 0 & 0 & 0 \\
 0 & 0 & 0 & \hskip-0.21cm-1 & 0 & 0 & 0 & 0 \\
 0 & 0 & 0 & 0 & 0 & 0 & 0 & 0 \\
 0 & 0 & 0 & 0 & 0 & \hskip-0.21cm-1 & 0 & 0 \\
 0 & 0 & 0 & 0 & 0 & 0 & \hskip-0.21cm-1 & 0 
\end{array} 
\end{array}}
 \right)\,, }
\end{equation}
where $\mathbf{1}$ stands for the $15\times15$ identity matrix while $\mathbf{0}$ always denotes suitable type of matrices with identically zero elements.     Taking then into account the decomposition 
\begin{equation}
\mathbb{A}^\mu \cdot \partial_\mu=\mathbb{A}^{\rm{D}} \cdot \rm{D} +
\mathbb{A}^{\Delta} \cdot \Delta + \mathbb{A}^{\delta} \cdot \delta +
\mathbb{A}^{\overline\delta} \cdot \overline\delta\,, 
\end{equation}
along with the explicit form of the derivative operators $\rm{D},\Delta,\delta,\overline\delta$, given in terms of the partial derivatives with respect to the Gaussian null coordinates $(u,r,x^3,x^4)$ in ${\mathcal{O}}$ by (\ref{dop}), we get that
\begin{eqnarray}
&&\mathbb{A}^u=\mathbb{A}^{\Delta}\\
&&\mathbb{A}^r=\mathbb{A}^{\rm{D}}+U\cdot\mathbb{A}^{\Delta}
  +\omega\cdot\mathbb{A}^{\delta}
  +\overline\omega\cdot\mathbb{A}^{\overline\delta}\\  
&&\mathbb{A}^A=X^A\cdot\mathbb{A}^{\Delta}
  +\xi^A\cdot\mathbb{A}^{\delta}
  +\overline\xi^A\cdot\mathbb{A}^{\overline\delta}\,.  
\end{eqnarray} 
In virtue of the explicit forms of the matrices $\mathbb{A}^{\rm{D}}$, $\mathbb{A}^{\Delta}$, $\mathbb{A}^{\delta}$ and $\mathbb{A}^{\overline\delta}$ given by (\ref{M1}) %, (\ref{M2}) 
and (\ref{M2}), it is straightforward to see then, that $\mathbb{A}^u$, $\mathbb{A}^r$ and $\mathbb{A}^A$ are Hermitian, i.e.,
\begin{equation}
\overline{\mathbb{A}}{}^\mu{}^T={\mathbb{A}}^\mu\,.
\end{equation}
Similarly, the combination $\mathbb{A}^\mu(\ell_\mu+n_\mu)$ is positive definite as  
\begin{equation}\label{detA}
\mathbb{A}^\mu(\ell_\mu+n_\mu)=\mathbb{A}^{\rm{D}}+\mathbb{A}^{\Delta} 
\end{equation}  
thereby its determinant, $det\left(\mathbb{A}^\mu(\ell_\mu+n_\mu)\right)$, takes the constant value $16$ throughout ${{\mathcal{O}}}$. 
  
\medskip

Consequently, the reduced Einstein-Maxwell equations, (EM1)-(EM21), comprise a first order symmetric hyperbolic system to which in the characteristic initial value problem in the smooth setting unique solutions are know to exist %\,\footnote{Note that the corresponding existence and uniqueness results are known to apply to real valued variables. To see that these results do also apply to the present case one needs to consider the system in which all the complex valued variables are replaced by their real and imaginary parts.} 
\cite{rendall} provided that all the transverse derivatives of the initial data $\mathbb{V}_0$ can be evaluated on the pair of null hypersurfaces $\mathcal{H}_1$ and $\mathcal{H}_2$.
In a great extent by adopting the argument of Friedrich in \cite{friedrich2} (see also \cite{kannar}) this requirement can be seen to hold, as follows. 
  
\medskip

Consider first the null hypersurface $\mathcal{H}_2$ where $\partial_r$ is transverse. As $\mathbb{V}_0$ is known the first order $\partial_r$ derivatives of the variables listed in $\mathbb{V}$---with the exception of that of $\Psi_0$ and $\phi_0$---can be evaluated algebraically on $\mathcal{H}_2$ by making use of (EM.1)-(EM.21). To evaluate the $\partial_r$-derivative of $\Psi_0$ and $\phi_0$ on $\mathcal{H}_2$ one needs to take the first order $\partial_r$-derivative of (EM.14) and (EM.19) which can be integrated along the generators of $\mathcal{H}_2$ as all the pertinent coefficients have already been determined and the initial data for $\partial_r \Psi_0$  and  $\partial_r \phi_0$ on $\mathcal{Z}$ is also known as $\Psi_0$ and $\phi_0$ comprise the reduced data on $\mathcal{H}_1$ and their $\partial_r$-derivatives are inner derivatives there. 

Consider now the system yielded by the modification of (EM.1)-(EM.21) such that (EM.14) and (EM.19) are replaced by their first order $\partial_r$-derivatives. The successive higher order $\partial_r$-derivatives of this system can be seen to comprise a set of ordinary differential equations containing coefficients and source terms which are algebraic in lower order derivative expressions already known on $\mathcal{H}_2$. Some of these equations can be solved algebraically, while the rest can be integrated, for the unknown derivatives, iteratively by making use of the initial data at $\mathcal{Z}$, which can always be evaluated there by making use of the inner derivatives on $\mathcal{H}_1$ as above. 

\medskip

On $\mathcal{H}_1$ where $\partial_u$ is transverse the procedure is similar although from the system (EM.1)-(EM.21) only the subsystem comprised by 
(EM.14)-(EM.17) and (EM.18)-(EM.20) can be used to determine algebraically the first order $\partial_u$-derivative of $\Psi_0, \Psi_1, \Psi_2, \Psi_3$ and $\phi_0, \phi_1$. However, $\partial_u$-derivatives of the complementary equations can be integrated, for the $\partial_u$-derivatives of the rest of the unknowns in $\mathbb{V}$, along the null generators of $\mathcal{H}_1$ as all the pertinent coefficients are already known and the initial data for the first order $\partial_u$-derivatives are determined on $\mathcal{Z}$ as the $\partial_u$-derivative is an inner derivative on $\mathcal{H}_2$. 

Again, by replacing the equations complementary to the ones (EM.14)-(EM.17) and (EM.18)-(EM.20) in the reduced Einstein-Maxwell equations  (EM.1)-(EM.21) with their $\partial_u$-derivatives we get a system of ordinary differential equations which can be solved step by step following an analogous procedure as applied on $\mathcal{H}_2$ above. 

According to the above argument all the $\partial_r$- and $\partial_u$-derivatives get to be determined on $\mathcal{H}_2$ and $\mathcal{H}_1$, respectively, which completes our proof.
{\hfill$\Box$}\bigskip 

Some remarks are in order now. Note first that by relation (\ref{detA}), along with $det\left(\mathbb{A}^\mu(\ell_\mu+n_\mu)\right)=16$ throughout ${{\mathcal{O}}}$, implies that the domain of  dependence $D[\mathcal{H}_1\cup\mathcal{H}_2]$ of the pair of null hypersurfaces, $\mathcal{H}_1$ and $\mathcal{H}_2$, in the smooth case, 
always extends as far as the Gaussian null coordinates are defined, i.e.\,the relation $\left\{J^+[\mathcal{Z}]\cup J^-[\mathcal{Z}]\right\}\cap \mathcal{O}\subset D[\mathcal{H}_1\cup\mathcal{H}_2]$ holds.

\bigskip 

Returning to the principal issues raised in \cite{CH} note first that whenever we have a solution $\mathbb{V}$ to the reduced Einstein-Maxwell equations, in virtue of the relations (\ref{m2}), (\ref{m3}) and (\ref{tet}), the functions $\xi^A,\omega,X^A,U$ uniquely determine the metric $g^{ab}$ and the frame vectors 
\begin{equation}
z_{00'}^a=\ell^a,\ \ z_{11'}^a=n^a,\ \ z_{01'}^a=m^a,\ \  z_{10'}^a=\overline{m}^{\,a}
\end{equation}
on $\mathcal{O}$, respectively. Similarly, in virtue of the relations in Table \ref{table:data0}, the spin-coefficients $\rho,\sigma,\tau,\alpha,\beta,\gamma,\lambda,$ $\mu,\nu$ determine a spin connection $\Gamma_{aa'bc}$ that can be associated with a connection $\nabla_{a}$, the action of which on a spinor ${w_d}^{e'}$, for example, is given as 
\begin{equation}\label{connection}
\nabla_{{ z}_{aa'}}{w_d}^{e'}={\bf z}_{aa'}({w_d}^{e'})-{{\Gamma_{aa'}}{}^c}{}_d\,{w_c}^{e'}+{{\overline{\Gamma}_{aa'}}{}^{e'}}{}_{c'}\,{w_d}^{c'}\,,  
\end{equation}
where the abbreviation ${\bf z}_{aa'}({w_d}^{e'})={{ z}^i_{aa'}}\frac{\partial {w_d}^{e'}}{\partial x^i}$ has been applied.   
Since $\Gamma_{aa'bc}=\Gamma_{aa'(bc)}$ this connection can be seen to be metric, i.e.\,$\nabla_ag_{bc}=0$. However, there is no guarantee that its torsion ${{t_{bb'}}^{aa'}}_{cc'}$ determined by the relation
\begin{eqnarray}\label{tors}
&&{{t_{bb'}}^{aa'}}_{cc'}\, {z}^i_{aa'}=\epsilon^{ae}\left[\Gamma_{bb'ec}\,{z}^i_{ac'}-\Gamma_{cc'eb}\,{z}^i_{ab'}\right]+ \epsilon^{a'e'}\left[\overline{\Gamma}_{bb'e'c'}\,\overline{z}^{\,i}_{ca'}-\overline{\Gamma}_{cc'e'b'}\overline{z}^{\,i}_{ba'}\right]\nonumber\\ && \phantom{{{t_{bb'}}^{aa'}}_{cc'}\, {e}^i_{aa'}= }-\left[{\bf z}_{cc'}({e}_{bb'}^i)-{\bf z}_{bb'}({z}_{cc'}^i)\right]
\end{eqnarray} 
vanishes. Finally, in virtue of the relations in Tables \ref{table:dataW} and \ref{table:dataM}, the Weyl and Maxwell spinor components $\Psi_0,\Psi_1,\Psi_2,\Psi_3,\Psi_4$ and $\phi_0,\phi_1,\phi_2$ determine a curvature and Maxwell tensors via the relations (\ref{riccimax}) and 
\begin{equation}\label{curv01}
R_{aa'bb'cc'dd'}=\epsilon_{a'b'}\,R_{abcc'dd'}+\epsilon_{ab}\,\overline{R}_{a'b'cc'dd'}\,, 
\end{equation}
where
\begin{equation}\label{curv0}
R_{abcc'dd'}=\left[\Psi_{abcd}+(\epsilon_{ac}\epsilon_{bd}+\epsilon_{ad}\epsilon_{bc})\, \Lambda \right]\epsilon_{c'd'}+k\,\phi_{ab}\,\overline{\phi}_{c'd'}\,\epsilon_{cd}\,, 
\end{equation}
and 
\begin{equation}\label{maxw}
F_{aba'b'}=\phi_{ab}\,\epsilon_{a'b'}+\overline{\phi}_{a'b'}\,\epsilon_{ab}\,, 
\end{equation}
respectively. Nevertheless, there is no guarantee that this curvature tensor coincides with that of the connection $\nabla_a$, and that all the source-free Maxwell equations 
\begin{equation}\label{maxNPeqs}
{\tf}_{a'a}={\nabla^f}_{a'}{\phi}_{fa}=0\,,
\end{equation}
are satisfied. 

\medskip

To have a clear picture on the background of the above implicit questions recall first that the curvature spinor $r_{abcc'dd'}$ determined by the spin connection $\Gamma_{aa'bc}$ can be given as 
\begin{eqnarray}\label{curv}
&&r_{abcc'dd'}=-{\bf z}_{dd'}(\Gamma_{cc'ab})+{\bf z}_{cc'}(\Gamma_{dd'ab})\nonumber \\&&\phantom{r_{abcc'dd'}=}-\epsilon^{ts}\left[\Gamma_{dd'at}\Gamma_{cc'sb}+ \Gamma_{td'ab}\Gamma_{cc'sd}-\Gamma_{cc'at}\Gamma_{dd'sb}- \Gamma_{tc'ab}\Gamma_{dd'sc}\right]\nonumber \\&&\phantom{r_{abcc'dd'}=}- \epsilon^{t's'}\left[\Gamma_{dt'ab}\overline{\Gamma}_{cc'sb}+ \Gamma_{cd'ab}\overline{\Gamma}_{dd's'c'}\right]+{{t_{cc'}}^{ss'}}_{dd'} \Gamma_{ss'ab}\,.
\end{eqnarray}
In addition, the curvature tensor 
\begin{equation}\label{curv0r}
r_{aa'bb'cc'dd'}=\epsilon_{a'b'}\,r_{abcc'dd'}+\epsilon_{ab}\,\overline{r}_{a'b'cc'dd'}\,
\end{equation}
satisfies the first and second Bianchi identities, which, with tensor indices, reads as
\begin{equation}\label{1stbianchi}
{r_{[jli]}}^k+\nabla_{[j}{t_l}{}^k{}_{i]}+{t_{[j}}{}^q{}_{l}{t_{i]}}{}^k{}_{q}=0\,,
\end{equation}
\begin{equation}\label{2ndbianchi}
{\nabla_{[j}\,r_{kl]i}}^q+t_{[j}{}^s{}_{k}\,{r_{l]is}}{}^q=0\,.
\end{equation}

It can be shown (see, e.g.\,Section 4.6 in \cite{penrose:rindler}) that the analogue of (\ref{1stbianchi}) holds for $R_{aa'bb'cc'dd'}$, with vanishing torsion, provided that the symmetry and reality assumptions $\Psi_{abcd}={\Psi}_{(abcd)}$, $\phi_{ab}={\phi}_{(ab)}$ and $\Lambda=\overline{\Lambda}$ are satisfied. It also follows from the results covered by  Sections 4.10 and 5.2 of \cite{penrose:rindler} that 
\begin{eqnarray}\label{2ndbianchiR}
&& \nabla_{ee'} R_{aa'bb'cc'dd'}+\nabla_{cc'} R_{aa'bb'dd'ee'}+\nabla_{dd'} R_{aa'bb'ee'cc'}=\\
&& \phantom{\nabla_{ee'} R+} \epsilon_{a'b'}\left[\epsilon_{e'd'}\,\epsilon_{dc}\,H_{c'abe} - \epsilon_{c'd'}\,\epsilon_{de}\,H_{c'abc} \right] + \epsilon_{ab}\left[\epsilon_{ed}\,\epsilon_{d'c'}\,\overline{H}_{ca'b'e'} - \epsilon_{cd}\,\epsilon_{d'e'}\,\overline{H}_{ca'b'c'} \right]\nonumber\,,
\end{eqnarray}
where is $H_{a'bcd}$ defined as 
\begin{eqnarray}\label{Hdef}
&& H_{a'bcd}={\nabla^f}_{a'}\Psi_{bcdf}-{\nabla^{f'}}_b \Phi_{cda'f'} \\ && \phantom{H_{a'bcd}}={\nabla^f}_{a'}\Psi_{bcdf}-k\left[\phi_{cd}\left({\nabla^{e'}}_b\overline{\phi}_{e'a'}\right)+\overline{\phi}_{e'a'}\left({\nabla_{b}}^{e'}\phi_{cd}\right)\right]\nonumber\,.
\end{eqnarray}
Note that in virtue of (\ref{2ndbianchiR}) the second Bianchi identity for $R_{aa'bb'cc'dd'}$---it has no torsion---can be seen to be equivalent to the vanishing of $H_{a'bcd}$. 

\medskip

Using the above relations the coupled Newman-Penrose and Maxwell equations, (NP.6.10.a-h), (NP.6.11.a-r), (NP.6.12.a-h) and (NP.A1.a-d),---in the specific gauge choice introduced in subsection \ref{gaugefix}---can be seen to be equivalent to the system 
\begin{equation}\label{NPeqs}
\begin{split}
{{t_{bb'}}^{aa'}}_{cc'} =0 \phantom{\,.} &  \\
r_{abcc'dd'}-R_{abcc'dd'}=0 \phantom{\,.} & \\
H_{a'bcd}=0 \phantom{\,.} & \\
{\tf}_{a'a}=0\,, &  
\end{split}
\end{equation}
respectively. Similarly, the reduced set of Einstein-Maxwell equations, (EM.1-4), (EM.5-13), (EM.14-18) and (EM.18-21), can be seen to be equivalent to
\begin{equation}\label{redNPeqs}
\begin{split}
{{t_{00'}}^{aa'}}_{cc'} =0 \phantom{\,.} & \\
r_{ab00'dd'}-R_{ab00'dd'}=0 \phantom{\,.} &  \\
H_{0'111}=0,\, \ H_{1'1bc}-H_{0'0bc}=0\,, \  H_{1'000}=0 \phantom{\,.}&  \\
{\tf}_{0'1}=0\,,\ {\tf}_{1'1}-{\tf}_{0'0}=0 \,,\ {\tf}_{1'0}=0\,, 
\end{split}
\end{equation}
respectively.

\medskip

It is not at all obvious that whenever $\mathbb{V}$ is a solution to the reduced set of field equations (\ref{redNPeqs}) it is also a solution to the full set of the coupled Newman-Penrose and Maxwell equations (\ref{NPeqs}). It is also important to emphasize that once we have an affirmative answer to this issue a clear characterization of the connection and the curvature is also yielded. More precisely, then, in virtue of (\ref{NPeqs}.a) and (\ref{NPeqs}.b), the metric, the connection and the curvature tensor determined by $\mathbb{V}$ are so that the connection is torsion free, and also the curvature that can be built up from the Weyl and Maxwell spinor components coincides with the curvature associated with the metric. 

\begin{theorem}\label{equivalent2} 
Consider an initial data specification $\mathbb{V}_0$, satisfying the inner Newman-Penrose and Maxwell equations on the initial data surface, and denote by $\mathbb{V}$ the associated unique solution to the reduced set of Einstein-Maxwell equations. Then, $\mathbb{V}$ is also a solution to the full set of the coupled Newman-Penrose and Maxwell equations. 
\end{theorem}
{\sl Proof:} The proof is given by a straightforward generalization of analogous arguments of Friedrich applied in \cite{friedrich,friedrich1}. Accordingly, in justifying that (\ref{NPeqs}) holds whenever (\ref{redNPeqs}) is satisfied we shall proceed as follows. 

Note first that since (\ref{redNPeqs}) guaranties that some of the equations in  (\ref{NPeqs}) are satisfied we need only to take care of the rest of the equations in (\ref{NPeqs}). By making use of these equations, it can be verified that the vanishing of the variables listed in 
\begin{equation}\label{variables}
 \tz=({{t_{bb'}}^{aa'}}_{cc'}, r_{abcc'dd'}-R_{abcc'dd'}, {H}_{0'000}, {H}_{0'100}, {H}_{0'110}, {\tf}_{0'0}\,; \,``\mathrm{complex\ \,conjugate}\,")
\end{equation}
guaranties that our assertion holds. 

We shall justify the vanishing of these variables by deriving a set of `subsidiary system' of differential equations which will be shown to be linear and homogeneous in the variables, and their complex conjugate, listed in $\tz$, and they comprise a strongly hyperbolic system. The justification of our assertion follows then by combining that these type of systems are know to possess identically zero solution for vanishing initial data, and that the vanishing of this data on $\mathcal{H}_1\cup\mathcal{H}_2$ is guaranteed by the assumption that $\mathbb{V}_0$ satisfies the inner Newman-Penrose and Maxwell equations on the initial data surface.

\medskip

Evolution equations for the torsion, ${{t_{bb'}}^{aa'}}_{cc'}$ and for the difference $r_{abcc'dd'}-R_{abcc'dd'}$ can be derived by writing down (\ref{1stbianchi}) and (\ref{2ndbianchi}) these identities for $j=1$, or in spinorial notation $j=00'$, and taking into account (\ref{redNPeqs}.a-b), along with the corresponding relations, with vanishing torsion, for $R_{aa'bb'cc'dd'}$, we obtain\,\footnote{
The explicit form of (\ref{homlin1}) and (\ref{homlin2}) can be derived by replacing in (2.5.5) and (2.5.6) of \cite{friedrich} the terms $e_i$, $e_{cc'}$, ${C^k}_{jli}$ and $C_{abcc'dd'}$ by $z_i$, $z_{cc'}$, ${R^k}_{jli}$ and $R_{abcc'dd'}$, respectively.}
\begin{equation}\label{homlin1} 
\nabla_{00'}\, {{t_{ll'}}^{kk'}}_{ii'}={{\mycal{T}_{ll'}}^{kk'}}_{ii'}\left({{t_{bb'}}^{aa'}}_{cc'},r_{abcc'dd'}-R_{abcc'dd'}\right)
\end{equation}
\begin{equation}\label{homlin2}
\nabla_{00'} [r_{iqkk'll'}-R_{iqkk'll'}] ={\mycal{R}_{iqkk'll'}}\left({{t_{bb'}}^{aa'}}_{cc'},r_{abcc'dd'}-R_{abcc'dd'}, {H}_{0'000}, {H}_{0'100}, {H}_{0'110}, {\tf}_{0'0} \right)
\end{equation}
where ${{\mycal{T}_{ll'}}^{kk'}}_{ii'}$ and ${\mycal{R}_{iqkk'll'}}$ are linear and homogeneous functions of their indicated variables, and their complex conjugate.

\medskip

By a completely analogous argument that was used to derive (2.5.7) in \cite{friedrich}, one can evaluate the contraction $\mycal{F}=\nabla_{aa'}{\tf}{\,}^{a'a}$ in two different ways. On the one hand side, we get\,\footnote{As, in virtue of (\ref{redNPeqs}.d), the spinor fields ${\tf}_{0'1}, {\tf}_{1'0}$ vanish identically the expressions $(\nabla_{10'}-z_{10'})\tf_{0'1}$ and $(\nabla_{01'}-z_{01'})\tf_{1'0}$ can be seen to be given as contractions of various components of $\tf_{a'a}$ with the connection spinor, i.e.,\,no derivatives of $\tf_{a'a}$ are involved on the right hand side of (\ref{mathcalF1}).} 
\begin{equation}\label{mathcalF1}
\left(\nabla_{00'}+\nabla_{11'}\right) {\tf}_{0'0}=\mycal{F}+(\nabla_{10'}-z_{10'}){\tf}_{0'1} +(\nabla_{01'}-z_{01'}){\tf}_{1'0}\,.
\end{equation}
On the other hand, by commuting covariant derivatives with torsion, the relation 
\begin{equation}\label{mathcalF2}
\mycal{F}=\left(r_{afe}{}^{s'f}{}_{s'}-R_{afe}{}^{s'f}{}_{s'} \right)\,\phi^{ea}+\frac12 \,\epsilon^{a'e'} {{t_{aa'}}^{bb'}}_{ee'}\nabla_{bb'} \phi^{ea} \,,
\end{equation}
can be derived, where the term $R_{afe}{}^{s'f}{}_{s'}$, which vanish based on the symmetry properties of the curvature spinor, was inserted. 

\medskip

In deriving evolution equations for ${H}_{0'000}, {H}_{0'100}, {H}_{0'110}$ it is advantageous to decompose $H_{a'bcd}$ by taking its irreducible parts 
\begin{eqnarray}\label{dec}
&& H_{a'bcd}=H_{a'(bc)d}-\frac13\,\epsilon_{bc}H_{a'}{}^e{}_{ed}-\frac13\,\epsilon_{bd}H_{a'}{}^e{}_{ec}-\frac13\,\epsilon_{cd}H_{a'b}{}^e{}_{e} \nonumber \\ && \phantom{H_{a'bcd}}=\mathcal{H}_{a'bcd}+\frac13\,\epsilon_{bc}\,\chi_{a'd}+\frac13\,\epsilon_{bd}\,\chi_{a'c}\,,
\end{eqnarray}
where $\mathcal{H}_{a'bcd}$ is the totally symmetric part of $H_{a'bcd}$, i.e.\,$\mathcal{H}_{a'bcd}=H_{a'(bcd)}\,,$ while
\begin{equation}\label{dec2}
\chi_{a'd}=H_{a'e}{}^e{}_d={\nabla^{ee'}} \Phi_{eda'e'}=k\,\left[\overline{\phi}{}_{a'e'}\,{\tf}{\,}^{e'}{}_{d}+\phi{}_{ed}\,\overline{\tf}{\,}^e{}_{a'}\right].
\end{equation}

With the help of this decomposition and an argument analogous to the one that was used to derive (2.5.7) in \cite{friedrich} the contraction $H_{bc}=\nabla_{aa'}H{}^{a'}{}_{bc}{}^a$ can be evaluated in two different ways. On the one hand side, we get 
\begin{eqnarray}\label{mathcal{H}}
&& \hskip-1.8cm \left(\nabla_{00'}+\nabla_{11'}\right) {H}_{0'110} - \nabla_{10'} {H}_{0'100}+\nabla_{00'}\chi_{0'1}-\nabla_{10'}\chi_{0'0}= H_{11} +(\nabla_{01'}-z_{01'}){H}_{0'111}\\
&& \hskip-1.8cm\left(\nabla_{00'}+\nabla_{11'}\right) {H}_{0'100}- \nabla_{01'} {H}_{0'110}-\nabla_{10'} {H}_{0'000}+\nabla_{00'}\chi_{0'0}=H_{10}\\ 
&& \hskip-1.8cm\left(\nabla_{00'}+\nabla_{11'}\right) {H}_{0'000}- \nabla_{01'}{H}_{0'100} -\nabla_{01'}\chi_{0'0} +\nabla_{00'}\chi_{1'0} =H_{00}+(\nabla_{10'} -z_{10'}){H}_{1'000}\,,  
\end{eqnarray}
while, on the other hand, by commuting covariant derivatives with torsion, the relation 
\begin{equation}\label{Hbc}
H_{bc}=\nabla_{a'(a}\nabla_{b)}{}^{a'} \Psi_{cd}{}^{ab}+\nabla_{a(a'}\nabla_{b')}{}^{a} \Phi_{cd}{}^{a'b'}-\nabla^{a'}{}_c\left[{\nabla^{ee'}} \Phi_{eda'e'}\right]\,.
\end{equation}
The first term in (\ref{Hbc}) is the correspondent of the expression ``$F_{cd}$'' below (2.5.7) of \cite{friedrich} with replacing $C_{abcc'dd'}$ by $R_{abcc'dd'}$, i.e.\,it reads as
\begin{equation}
\nabla_{a'(a}\nabla_{b)}{}^{a'} \Psi_{cd}{}^{ab}=2\,\Psi_{t(dab}\left[{r}^{ab}{}_{c)}{}^{s't}{}_{s'}- {R}^{ab}{}_{c)}{}^{s't}{}_{s'}\right] - \frac12\,\epsilon^{a'b'}\,{{t_{aa'}}^{ff'}}_{bb'}\nabla_{ff'}\Psi_{cd}{}^{ab}\,.
\end{equation}
The middle term (\ref{Hbc}) can be given as 
\begin{eqnarray}\label{middle}
&&\nabla_{a(a'}\nabla_{b')}{}^{a} \Phi_{cd}{}^{a'b'}= -\left[\overline{r}_{a'b'(c|s'}{}^{qs'}-\overline{R}_{a'b'(c|s'}{}^{qs'}\right] \Phi_{q|d)}{}^{a'b'}\\ &&\phantom{\nabla_{a(a'}\nabla_{b')}{}^{a} \Phi_{cd}{}^{a'b'}=}+ \left[\overline{r}_{a'b'sq'}{}^{s(a'|}-\overline{R}_{a'b'sq'}{}^{s(a'|}\right] \Phi_{cd}{}^{q'|b')}  -\frac12\,\epsilon^{st}\,{{t_{sa'}}^{ff'}}_{tb'}\nabla_{ff'}\Phi_{cd}{}^{a'b'}\,,\nonumber
\end{eqnarray}
where the identically zero terms $\overline{R}_{a'b'(c|s'}{}^{qs'}\, \Phi_{q|d)}{}^{a'b'}=0$ and $\overline{R}_{a'b'sq'}{}^{s(a'|}\, \Phi_{cd}{}^{q'|b')}=0$---these relations hold based on the symmetry properties of $R_{afe}{}^{s'f}{}_{s'}$---were inserted.

Concerning the third term in (\ref{Hbc}) note that by (\ref{dec2}) 
\begin{equation}\label{dec3}
\nabla^{a'}{}_c\left[{\nabla^{ee'}} \Phi_{eda'e'}\right]=\nabla^{a'}{}_c \chi_{a'd}=\nabla_{c0'} \chi_{1'd}-\nabla_{c1'} \chi_{0'd}\,.
\end{equation}

By combining all the relations (\ref{dec2})-(\ref{dec3}) we infer that the variables ${H}_{0'000}, {H}_{0'100}, {H}_{0'110}$ are subject to the evolution equations 
\begin{eqnarray}
&& \hskip-1.8cm \left(\nabla_{00'}+\nabla_{11'}\right) {H}_{0'110} - \nabla_{10'} {H}_{0'100} = \mycal{H}_{0'110} \label{Heq1}\\ 
&& \hskip-1.8cm\left(\nabla_{00'}+\nabla_{11'}\right) {H}_{0'100}- \nabla_{01'} {H}_{0'110}-\nabla_{10'} {H}_{0'000}+k\left[\overline{\phi}{}_{1'1'}\,\nabla_{01'}{\tf}{\,}{}_{0'0}+\phi{}_{00}\,\nabla_{01'}\overline{\tf}{\,}{}_{0'0}\right]
= \mycal{H}_{0'100}\\ 
&& \hskip-1.8cm\left(\nabla_{00'}+\nabla_{11'}\right) {H}_{0'000}- \nabla_{01'} {H}_{0'100}= \mycal{H}_{0'000}\,, \label{Heq2}
\end{eqnarray}
where $\mycal{H}_{0'110}, \mycal{H}_{0'100}, \mycal{H}_{0'000}$ are linear and homogeneous functions of the variables listed in (\ref{variables}), respectively. 

\medskip

Note finally that the subsidiary equations, (\ref{homlin1}), (\ref{homlin2}),  (\ref{mathcalF1}),  (\ref{mathcalF2}), (\ref{Heq1}) - (\ref{Heq2}), along with their complex conjugate, can be seen to comprise a linear homogeneous strongly hyperbolic system \cite{reula}\,\footnote{A clear characterization of strongly hyperbolic systems can be find. e.g.\,in \cite{reula}.}$^,$\footnote{Note that by replacing ${H}_{0'110}$ by the combination ${H}_{0'110}+k\left[\overline{\phi}{}_{1'1'}\,\nabla_{01'}{\tf}{\,}{}_{0'0}+\phi{}_{00}\,\nabla_{01'}\overline{\tf}{\,}{}_{0'0}\right]$ and by applying the combination of (\ref{mathcalF1}) and  (\ref{mathcalF2}), the yielded system evolution equations can be seen to be a symmetric hyperbolic one for the pertinent set of variables.} for the unknowns and their complex conjugate listed in (\ref{variables}). Since the assumption that $\mathbb{V}_0$ satisfies the inner Newman-Penrose and Maxwell equations on the initial data surface guaranties that all of these 
variables vanish\,\footnote{The check of this is left to the readers.} on $\mathcal{H}_1\cup\mathcal{H}_2$ they are identically zero everywhere in the domain of dependence of $\mathcal{H}_1\cup\mathcal{H}_2$, which, in turn, justifies our assertion. 
{\hfill$\Box$}\bigskip

\section{Appendix}\label{Appendix B}

In this section some of the basic relations of the Newman-Penrose 
formalism---specific to the gauge choices that have been made in Subsection\,\ref{gaugefix} are recalled.

\bigskip

\renewcommand{\theequation}{$NP.6.10.$}
{\it The metric equations:}
\begin{subequations}\label{NP.6.10}
\begin{align}
\mathrm{D}\xi ^A=&\,{\rho }\,\xi ^A+{\sigma }\,\bar\xi^A \\ 
\mathrm{D}{\omega }=&\,{\rho\, \omega }+\sigma\, \overline{\omega }-\tau  \\ 
\mathrm{D}X^A=&\,{\tau }\,\bar\xi^A+{\bar{\tau}\,\xi }^A \\ 
\mathrm{D}U=&\,{\tau }\,\overline{\omega }+\bar{\tau}\,\omega -(\gamma +
\overline{\gamma }) \\ 
{\delta }X^A-{\Delta }{\xi }^A=&\,({\mu }+{\bar{\gamma}}-\gamma )\,{\xi }
^A+{\bar{\lambda}}\,\bar\xi^A \\ 
{\delta }\overline{\xi }^A-\overline{\delta }{\xi }^A=&\,(\overline{\beta }
-\alpha )\,{\xi }^A+(\overline{\alpha }-\beta )\,\bar\xi^A \\ 
{\delta }\overline{\omega }-\overline{\delta }\omega =&\,(\overline{\beta }
-\alpha )\,\omega +(\overline{\alpha }-\beta )\,\overline{\omega }+(\mu -
\overline{\mu }) \\ 
{\delta }U-{\Delta }\omega=&\,({\mu }+{\bar{\gamma}}-\gamma
)\,\omega +{\bar{\lambda}}\,\overline{\omega }-\overline{\nu }
\end{align}
\end{subequations}

Note that these equations, in the applied specific gauge, can be see to be equivalent to the vanishing of the torsion tensor associated with the covariant derivative operator associated with the complex null tetrad $\{\ell^a,n^a,m^a,\overline{m}^a\}$.  

%\newpage

\bigskip

\renewcommand{\theequation}{$NP.6.11.$}
{\it The Ricci identities:} 
\begin{subequations}\label{NP.6.11}
\begin{align}
\mathrm{D}{\rho }=&\,{\rho }^2+{\sigma\, \bar{\sigma}}+{{\Phi }_{00}} \\
\mathrm{D}{\sigma }=&\,2\,{\rho\, \sigma }+{{\Psi }_0}  \\
\mathrm{D}{\tau }=&\,{\tau\, \rho }+{\bar{\tau}\,\sigma }+{{\Psi }_1}+{{\Phi }
_{01}} \\ 
\mathrm{D}{\alpha }=&\,{\rho \,\alpha }+{\beta\, \bar{\sigma}}+{{\Phi }_{10}} \\
\mathrm{D}\beta =&\,{\alpha \,\sigma }+{\rho\, \beta }+{{\Psi }_1} \\
\mathrm{D}{\gamma }=&\,{\tau\, \alpha }+{\bar{\tau}\,\beta }+{{\Psi }_2}-{\Lambda 
}+{{\Phi }_{11}} \\
\mathrm{D}{\lambda }=&\,{\rho\, \lambda }+{\bar{\sigma}\,\mu }+{{\Phi }_{20}} \\ 
\mathrm{D}{\mu }=&\,{\rho \,\mu }+{\sigma\,\lambda }+{{\Psi }_2}+2\,{\Lambda }\\ 
\mathrm{D}{\nu}=&\,{\bar{\tau}\,\mu}+{\tau\,\lambda}+{{\Psi}_3}+{{\Phi}_{21}} \\ 
{\Delta }{\lambda }-{\bar{\delta}}{\nu }=&\,({\bar{\gamma}}-3\,{\gamma }-{\mu 
}-{\bar{\mu}})\,{\lambda }+(3\,{\alpha }+{\bar{\beta}}-{\bar{\tau}})\,{\nu }-{{
\Psi }_4} \\ 
{\delta }{\rho }-{\bar{\delta}}{\sigma }=&\,({\bar{\alpha}}+{\beta})\,{\rho}
-(3{\alpha }-{\bar{\beta}})\,{\sigma }-{{\Psi }_1}+{{\Phi }_{01}} \\
{\delta }{\alpha }-{\bar{\delta}}{\beta }=&\,{\rho \,\mu }-{\sigma\, \lambda}+
{\alpha \,\bar{\alpha}}+{\beta \,\bar{\beta}}-2\,{\alpha \,\beta }-{{\Psi }_2}+
{\Lambda }+{{\Phi }_{11}} \\
{\delta }{\lambda }-{\bar{\delta}}{\mu }=&\,({\alpha }+{\bar{\beta}})\,{\mu}
+({\bar{\alpha}}-3\,{\beta })\,{\lambda }-{{\Psi }_3}+{{\Phi }_{21}} \\ 
\delta{\nu}-{\Delta}{\mu}=&\,{\mu}^2+{\lambda \,\bar{\lambda}}+({\gamma}
+{\bar{\gamma}})\,{\mu}+({\tau}-{\bar{\alpha}}-3\,{\beta})\,{\nu}+{{\Phi}_{22}} \\ 
{\delta }{\gamma }-{\Delta }{\beta }=&\,{\mu\, \tau }-{\sigma\, \nu }-({\gamma }
-{\bar{\gamma}}-{\mu })\,{\beta }+{\alpha\, \bar{\lambda}}+{{\Phi }_{12}} \\
{\delta }{\tau }-{\Delta }{\sigma }=&\,{\mu\, \sigma }+{\bar{\lambda}\,\rho}+({
\tau }-{\bar{\alpha}}+{\beta })\,{\tau }-(3\,{\gamma }-{\bar{\gamma}})\,{\sigma }+{
{\Phi }_{02}} \\ 
{\Delta }{\rho }-{\bar{\delta}}{\tau }=&\,-{\rho\, \bar{\mu}}-{\sigma\, \lambda 
}+({\gamma }+{\bar{\gamma}})\,{\rho }-({\bar{\tau}}+{\alpha }-{\bar{\beta}})
\,{\tau }-{{\Psi }_2}-2\,{\Lambda } \\
{\Delta }{\alpha }-{\bar{\delta}}{\gamma }=&\,{\rho\, \nu}-({\tau }+{\beta })\,
{\lambda }+({\bar{\gamma}}-{\bar{\mu}})\,{\alpha }+({\bar{\beta}}-{\bar{\tau}})\,
{\gamma }-{{\Psi }_3}
\end{align}
\end{subequations}

\renewcommand{\theequation}{$NP.6.12.$}
{\it The Bianchi identities:}
\begin{subequations}\label{NP.6.12}
\begin{align}
\mathrm{D}[\Psi _1{-{\Phi }_{01}}]-\overline{\delta }\Psi _0+\delta {\Phi}_{00}=&\,-4\,\alpha\,\Psi _0+4\,\rho\,\Psi _1+2\tau\,{\Phi }_{00}-2\,\rho\,{\Phi}_{01}
-2\,\sigma \,{\Phi }_{10} \\ 
\mathrm{D}[\Psi {_2+2\,\Lambda ]-}\overline{\delta }[\Psi _1+{{\Phi }_{01}}]
+\Delta {{\Phi }_{00}}=&\, -\lambda\, \Psi _0-2\,\alpha\, \Psi _1+3\,\rho\, \Psi _2 +(2\,\gamma +2\,\overline{\gamma }-\overline{\mu })\,{{\Phi }_{00}}\\ &\, 
{-2\,(\alpha +}\overline{\tau })\,{{\Phi }_{01}} -{2\,}{\tau\, {{\Phi }_{10}}+2\,\rho\,{\Phi }_{11}+}\overline{\sigma }\,{{\Phi }_{02}}
\nonumber \\
\mathrm{D}[\Psi _3{-{\Phi }_{21}}]-\overline{\delta }[\Psi _2+2\,\Lambda]+\delta {\Phi }_{20}=&\,-2\,\lambda\, \Psi _1+2\,\rho \,\Psi _3+2\,\mu\, {{\Phi }}
_{10}-2\,{(\beta }-\overline{\alpha })\,{{\Phi }_{20}}\\ &\,
-{2\,\rho {\Phi }_{21}}\nonumber \\
\mathrm{D}\Psi _4-\overline{\delta }[\Psi _3+{{\Phi }_{21}}]+\Delta {{
\Phi }_{20}}=&\,
-3\,\lambda \,\Psi _2+2\,\alpha\, \Psi _3+\rho \,\Psi _4+2\,\nu\, {{\Phi }
}_{10}-2\,\lambda \,{\Phi }_{11} \\ &\,
-{(}2\,\gamma -2\,\overline{\gamma }+\overline{\mu })\,{{\Phi }_{20}}
{-2\,(\overline{\tau }-\alpha )\,{\Phi }_{21}+}\overline{\sigma }\,{{\Phi }_{22}}
\nonumber \\ 
\Delta\Psi_0-\delta[\Psi _1{+{\Phi}_{01}}]+\mathrm{D}{{\Phi }_{02}}
=&\, (4\,\gamma -\mu )\,\Psi _0-2\,(2\,\tau+\beta )\,\Psi_1+3\,\sigma\,\Psi_2 -{\bar{\lambda}}\,{\Phi }_{00} \\ &\,
-2\,\beta\,{\Phi }_{01} +2\,\sigma\,{\Phi}_{11}+\rho\,{\Phi }_{02} \nonumber\\
\Delta [\Psi _1-{{\Phi }_{01}}]-\delta [\Psi _2+2\,{\Lambda }]+\overline{
\delta }{{\Phi }_{02}}=&\,
\nu \,\Psi _0+2\,(\gamma -\mu )\,\Psi _1-3\,\tau\, \Psi _2+2\,\sigma\, \Psi_3-
\overline{\nu }\,{\Phi }_{00}  \\ &\,
+2\,(\overline{\mu }{-\gamma )\,{\Phi }_{01}+(3\,\alpha }
-\overline{\beta })\,{{\Phi }_{02}+2\,\tau \,{\Phi }_{11}-2\,\rho\,{\Phi }_{12}}
\nonumber\\ 
\Delta[\Psi_2+2\,\Lambda]-\delta[\Psi _3+{{\Phi}_{21}}]+\mathrm{D}{\Phi }_{22}
=&\, 2\,\nu\,\Psi _1-3\,\mu\, \Psi _2-2\,{\overline{\alpha }}\,\Psi _3+\sigma \,\Psi
_4 -2\,\mu \,{{\Phi }_{11}} \\ &\,
-\overline{\lambda }\,{\Phi }_{20}+{2\,\beta\,{\Phi }_{21}}+\rho\,{\Phi }_{22}\nonumber \\
\Delta [\Psi _3{-{\Phi }_{21}}]-\delta \Psi _4+\overline{\delta }{{\Phi }
_{22}}=&\,  
3\,\nu\, \Psi _2-2\,(\gamma +2\,\mu )\,\Psi _3+(4\,\beta -\tau)\,\Psi _4 -2\,\nu\, {{\Phi }_{11}}
\\ &\, -\overline{\nu }\,{{\Phi }_{20}} +{2\,\lambda\, {\Phi }_{12}+2\,(\gamma +
\overline{\mu })\,{{\Phi }_{21}-}}\overline{\tau }\,{{{\Phi }_{22}}}\nonumber 
\end{align}
\end{subequations}

\renewcommand{\theequation}{NP.A1.}
{\it The Maxwell equations:}
\begin{subequations}\label{A1}
\begin{align}
\mathrm{D}\phi_1 - \overline{\delta } \phi_0 =&\, - 2 \,\alpha \,\phi_0 + 2\, \rho\, \phi_1 \\
\mathrm{D}\phi_2 - \overline{\delta } \phi_1 =&\, - \lambda\, \phi_0 + \rho\, \phi_2 \\
\delta\phi_1 - \Delta \phi_0 =&\, (\mu -2 \gamma)\, \phi_0 + 2\, \tau \,\phi_1 - \sigma\, \phi_2 \\
\delta\phi_2 - \Delta \phi_1 =&\, -\nu\, \phi_0 + 2\, \mu\, \phi_1 + (\tau-2\beta)\, \phi_2\,.
\end{align}
\end{subequations}
%%%%%%%%%%%%%%%%%%%%%%%%%%%%%%%%%%%%%%%%%%%%%%%%%%%%%%%%%%%%%%%%%%%%

\end{document}